\documentclass[journal,onecolumn]{IEEEtran}
\usepackage{cite,amsmath,amssymb,amsthm,mathtools,mathrsfs}
\usepackage{enumitem,booktabs,array,tabularx,makecell,multirow,microtype}
\usepackage{stmaryrd}
\usepackage{color}
\usepackage{pgfplots}
\usepackage[colorlinks=true,linkcolor=red,citecolor=blue,urlcolor=blue]{hyperref}

\hypersetup{pdftitle={Optimal Impure (r,delta)-Quantum Locally Recoverable Codes with Explicit Minimum Localities},
 pdfauthor={Yang Li, Shitao Li, Gaojun Luo, San Ling}}
\theoremstyle{definition}
\newtheorem{theorem}{Theorem}
\newtheorem{problem}[theorem]{Problem}
\newtheorem{lemma}[theorem]{Lemma}
\newtheorem{corollary}[theorem]{Corollary}
\newtheorem{remark}[theorem]{Remark}

\newtheorem{definition}[theorem]{Definition}
\newcommand{\F}{\mathbb{F}}
\newcommand{\CSS}{\mathrm{CSS}}
\newcommand{\wt}{\mathrm{wt}}
\newcommand{\supp}{\operatorname{supp}}
\newcommand{\Span}{\operatorname{span}}
\newcommand{\C}{\mathcal{C}}
\newcommand{\D}{\mathcal{D}}
\newcommand{\diag}{\operatorname{diag}}
\newcommand{\ev}{\operatorname{ev}}
\newcommand\dsb[1]{\llbracket #1 \rrbracket}
\newcommand\Dsb[1]{\left\llbracket #1 \right\rrbracket}

\definecolor{movedgreen}{rgb}{0,0.42,0.22}

\definecolor{deletedgray}{gray}{0.42}

\newcommand{\theme}[1]{#1}

\title{Optimal Impure $(r,\delta)$-Quantum Locally Recoverable Codes
\theme{with Explicit Minimum Localities}}
\author{Yang Li, Shitao Li, Gaojun Luo, San Ling
\thanks{This research was supported by Nanyang Technological University under
Research Grant 04INS000047C230GRT01.}
\thanks{Yang Li is with the School of Physical and Mathematical
Sciences, Nanyang Technological University, 21 Nanyang Link, Singapore 637371,
Singapore (e-mail: yanglimath@163.com).}
\thanks{Shitao Li is with the School of Internet, Anhui University, Hefei, Anhui 230039, China (e-mail: lishitao0216@163.com).}
\thanks{Gaojun Luo is with the School of Mathematics, Nanjing University of
Aeronautics and Astronautics, Nanjing 210016, China (e-mail: gaojun.luo@nuaa.edu.cn).}
\thanks{San Ling is with the School of Physical and Mathematical
Sciences, Nanyang Technological University, 21 Nanyang Link, Singapore 637371, Singapore (e-mail: lingsan@ntu.edu.sg).
He is also with VinUniversity, Vinhomes Ocean Park, Gia Lam,
Hanoi 100000, Vietnam (e-mail: ling.s@vinuni.edu.vn).}
}

\begin{document}
\maketitle
\begin{abstract}
Quantum locally recoverable codes with $(r,\delta)$-locality ($(r,\delta)$-qLRCs)
are quantum error-correcting codes in which $\delta-1$ erased qudits can be recovered by accessing only $r$ other qudits.
The minimum locality of a qLRC is the least $r$ such that it has $(r,\delta)$-locality.
This paper aims to construct optimal impure $(r,\delta)$-qLRCs with disjoint recovering sets and explicit minimum localities,
where the optimality is measured by the Singleton-like bound of
Guruswami, Kshirsagar, and Trivedi (GKT-optimal).
Such qLRCs have particular advantages in recovering as many quantum erasures as possible
while operating in parallel and with minimal qudit access overhead.

We first characterize the recovering sets and the minimum locality of any CSS code by local relative distances,
which are relative minimum distances of punctured codes with respect to shortened codes.
We then construct impure qLRCs as CSS codes from nested pairs of generalized monomial-Cartesian codes.
Grouping the monomials by their degree in the last variable, we reduce both their minimum distances and their local recovery
to nested pairs of codes in the remaining variables,
and the coordinates with the same values of these variables form disjoint recovering sets.
In this way, we obtain four infinite families of GKT-optimal impure $(r,\delta)$-qLRCs
with disjoint recovering sets, and determine their minimum localities
and the recovering sets attaining them for every admissible $\delta$,
except for a partial range of $\delta$ in one family.
These minimum localities need not be monotone in $\delta$.
In particular, two of the families contain GKT-optimal impure subfamilies of CSS codes
$\CSS(\C,\C)$ with $(r,\delta)$-locality for some $r<n-\delta+1$
at values of $\delta$ for which $\C$ has no classical $(r',\delta)$-locality for any $0<r'<n$,
where $n$ is the length of $\C$.

\end{abstract}

\section{Introduction}\label{sec:introduction}
Throughout this paper, we let $\F_q$ denote the finite field of size $q$.
Let $\mathbb{C}$ be the field of complex numbers and let $\mathbb{C}^q$ be the $q$-dimensional Hilbert space over $\mathbb{C}$.
An $\dsb{n,\kappa,d}_q$ quantum code
is a $q^\kappa$-dimensional subspace of $(\mathbb C^q)^{\otimes n}$ that corrects the erasure of any
$d-1$ qudits, where erasure in quantum error correction
refers to an error whose position in a codeword is known \cite{GBP1997}.
Given a classical linear code $\C$ over $\F_q$, we denote its dual code by $\C^\perp$.
The well-known CSS construction \cite{CS1996,S1996,KKKS2006} provides an efficient method for constructing
a quantum stabilizer code $\CSS(\C_X,\C_Z)$ from two classical linear codes $\C_X$ and $\C_Z$ satisfying
the dual-containing condition $\C_Z^\perp\subseteq\C_X$, whose minimum distance $d$ is the smaller of
the {\em relative minimum distances} $d(\C_X\setminus\C_Z^\perp)$ and
$d(\C_Z\setminus\C_X^\perp)$ \cite{KKKS2006}.
We also call $\CSS(\C_X,\C_Z)$ a {\em CSS code},
and it is {\em impure} if its minimum nonidentity stabilizer weight
$\sigma=\min\{d(\C_X^\perp),~d(\C_Z^\perp)\}$ satisfies $\sigma<d$ \cite{KKKS2006},
or equivalently, if $d>\min\{d(\C_X),~d(\C_Z)\}$;
and {\em pure} otherwise \cite{R1999}.

\subsection{Quantum locally recoverable codes}\label{subsec:qlrc}

Recall that classical locally recoverable codes (cLRCs) were introduced in \cite{GHSY2012} for
distributed storage, where a code has locality $r$ if every code symbol can be recovered
from at most $r$ other symbols. Prakash \emph{et al.}~\cite{PKLK2012} extended this to
$(r,\delta)$-locality, where every coordinate lies in a recovering set of size at most
$r+\delta-1$ on which the punctured code has minimum distance at least $\delta$.
In particular, $(r,2)$-cLRCs coincide with cLRCs with locality $r$.

As a quantum analogue, quantum locally recoverable codes (qLRCs) were introduced
for large-scale quantum data storage by Golowich and Guruswami \cite{GG2025}.
The term ``locally recoverable'' in the quantum context means that any single erased qudit can be recovered by a
recovery channel acting on at most $r$ other qudits.
To correct multiple erasures simultaneously, Galindo \emph{et al.}~\cite{GHMM2026FFA} extended the concept of $r$-locality
to $(r,\delta)$-locality, which is recalled in Definition~\ref{def:quantum-locality} and was further revisited
as special cases of hierarchical qLRCs in \cite{GKT2026}.
They also characterized the local recoverability of CSS codes
by shortened and punctured constituent codes.
Roughly speaking, a quantum code has $(r,\delta)$-locality if every qudit lies in
an $(r,\delta)$-recovering set $J$ with $\delta\le|J|\le r+\delta-1$ such that
any $\delta-1$ erased qudits in $J$ can be recovered by a recovery channel acting only on $J$,
that is, from the remaining at most $r$ qudits of $J$.
Matsumoto \cite{M2025} further showed that quantum local recovery reduces
the number of measurements in quantum erasure correction.
Motivated by these applications, qLRCs have attracted considerable attention, 
where optimal constructions were given in \cite{AG2026,LCEL2025,LLLLL2026,XZS2025,CZ2026,ZC2026}
and bounds were derived in \cite{LLLL2025,SB2026,KB2026bound,LLLLZ2026,LJX2026}. 
More related works can also be found in the references therein.

\subsection{The GKT-optimality and minimum locality of qLRCs}\label{subsec:gkt-minloc}

The works listed above mainly focus on pure qLRCs,
and the study of impure qLRCs is still in its infancy.
This is also reflected in the known bounds.
Although several bounds \cite{GG2025,LLLLL2026,LCEL2025,BGL2025} apply to impure codes
for $\delta=2$,
all the bounds for $(r,\delta)$-qLRCs with $\delta\ge3$ in
\cite{GHMM2026FFA,SB2026,KB2026bound} hold only for pure codes,
and finding efficient bounds for general $(r,\delta)$-qLRCs with $\delta\ge3$ was
raised as an open problem in \cite{GHMM2026QIP}.
To the best of our knowledge, the bound of Guruswami, Kshirsagar, and Trivedi
\cite[Theorem~2.10]{GKT2026} is the first one that also applies
to impure qLRCs with $\delta\ge3$, under the assumption of disjoint recovering sets.
See Definition~\ref{def:recovering-sets} for the exact meaning of disjoint recovering sets.
Note also that Ma and Shi \cite{MS2026} subsequently extended the scope of such a bound to
arbitrarily overlapping recovering sets.
Their general bound has a different expression,
and the GKT bound can be strictly stronger when the disjointness condition holds.
The scarcity of such bounds and the additional restriction on the recovering sets explain why, to date,
few infinite families of impure $(r,\delta)$-qLRCs are known to be optimal with respect to
the bound of Guruswami \emph{et al.} \cite{GKT2026}.

We call the bound of Guruswami, Kshirsagar, and Trivedi \cite[Theorem~2.10]{GKT2026}
the \emph{GKT bound} in this paper. For an $\dsb{n,\kappa,d}_q$ quantum code with disjoint $(r,\delta)$-recovering sets,
the GKT bound states that
\begin{equation}\label{eq:gkt-endpoint}
 \kappa\le\left(1-\frac{2(\delta-1)}{r+\delta-1}\right)n
 -2\left(d-1-(\delta-1)
           \left\lceil\frac{d-1}{r}\right\rceil\right).
\end{equation}
Since the right-hand side of \eqref{eq:gkt-endpoint} is not monotone in $d$ when $\delta\ge3$,
we say that an $\dsb{n,\kappa,d}_q$ qLRC with disjoint $(r,\delta)$-recovering sets is
\emph{GKT-optimal} if \eqref{eq:gkt-endpoint} holds for its $n,~\kappa,~d,~r,~\delta$
but fails when $d$ is replaced by $d+1$ with other parameters fixed.

By Definition~\ref{def:quantum-locality}, a code with $(r,\delta)$-locality naturally
has $(r',\delta)$-locality for every $r'\ge r$.
Hence, the locality $r$ stated for the CSS construction is only an upper bound
on the number of qudits accessed in local recovery \cite{LLLLZ2026}.
From the application perspective, this number should be as small as possible.
Therefore, given an $(r,\delta)$-qLRC $\mathcal Q$, we generally prefer to have a smaller $r$ and a larger $\delta$.
In other words, for {\em fixed} $\delta$, it is important to determine the {\em smallest} $r$ for which a qLRC has $(r,\delta)$-locality.
We call this $r$ the \emph{minimum locality} $r_{\mathcal Q}^{\min}(\delta)$ of the qLRC $\mathcal Q$,
and define the minimum locality $r_{\C}^{\min}(\delta)$ of a linear code $\C$ in the same way.
It measures the number of qudits that have to be accessed, in the worst case over all qudits, to locally repair $\delta-1$ erasures.

\subsection{Our motivations for constructing GKT-optimal impure qLRCs with explicit minimum localities}\label{subsec:motivation}

Compared to plenty of optimal pure qLRCs, the lack of optimal constructions of
infinite families of impure $(r,\delta)$-qLRCs provides the first motivation.
In addition, the following three observations motivate our study.

\begin{enumerate}
\item {\em Impure CSS codes may have minimum localities unavailable to their constituent codes.}
In \cite[Sec.~4]{GHMM2026QIP}, Galindo \emph{et al.} speculated that the locality of an impure code $\mathcal{Q}=\CSS(\C,\C)$ might differ from that of $\C$.
For fixed $\delta\ge2$, there are two possibilities:
\begin{itemize}
  \item $r_{\mathcal Q}^{\min}(\delta) < r_{\C}^{\min}(\delta)$ when both $\mathcal Q$ and $\C$ have $(r,\delta)$-locality, or
  \item $\mathcal Q$ has quantum $(r,\delta)$-locality
  while $\C$ has no classical $(r',\delta)$-locality for any $0< r'<n$.
\end{itemize}
\hspace{4mm} If the constituent code $\C$ has classical $(r',\delta)$-locality for some $r'$,
it follows from $\C^{\perp}\subseteq \C$
that $\delta\le d(\C)\le d(\C^\perp)$.
Under this condition, the first possibility was ruled out by \cite[Theorem~29]{GHMM2026FFA},
where the authors showed that
$\CSS(\C,\C)$ has quantum $(r,\delta)$-locality if and only if $\C$ has classical $(r,\delta)$-locality.
In addition, for a pure CSS code, the condition $\delta\le d(\C^\perp)$ is also automatically satisfied for every admissible recovery parameter $\delta\le d$,
which implies that the impure requirement is also necessary for the second possibility to occur.
Note also from Definition~\ref{def:quantum-locality} that every quantum code with minimum distance $d\ge\delta$ naturally has $(n-\delta+1,\delta)$-locality.
We therefore consider the second possibility
and derive the following problem.

\medskip

\begin{problem}\label{prob:quantum-classical-locality}
Let $\C$ be a linear code of length $n$ satisfying $\C^\perp\subseteq \C$.
Can an impure CSS code $\CSS(\C,\C)$ have $(r,\delta)$-locality for some $0< r< n-\delta+1$,
while $\C$ has no classical $(r',\delta)$-locality for any $0<r'<n$?
\end{problem}
\medskip

\item {\em The GKT bound is sharper and disjoint recovering sets are practical.}
As mentioned above, the GKT bound exploits the disjointness of recovering sets
and can be strictly stronger than the bounds for arbitrary recovering sets established in \cite{MS2026}.
A sharper bound is closer to the achievable parameters, and hence 
it is more likely to be attained by explicit qLRCs.
Disjoint recovering sets also have practical advantages.
Recovery channels on disjoint recovering sets act on disjoint qudits and can be applied in parallel,
whereas a qudit shared by two recovering sets might not be used by both recovery channels at the same time,
since an unknown quantum state cannot be cloned \cite{WZ1982}.
Moreover, since measurements disturb the measured qudits,
disjointness ensures that each qudit is measured only in the recovery of a single recovering set.
Therefore, it is more feasible to construct codes attaining the GKT bound
and such GKT-optimal codes seem to be more convenient for quantum local recovery.

\item {\em Impurity can increase the erasure-correction capability.}
Galindo \emph{et al.}~\cite{GHMM2026QIP} constructed a general family of impure qLRCs 
with several explicit subfamilies and
showed that impure codes can exceed the pure bounds \cite{LLLL2025} for quantum local recovery.
It implies that impure codes may correct more erasures than pure codes with the same length, dimension, and locality.
However, the structure of recovering sets of the impure qLRCs is not discussed in \cite{GHMM2026QIP},
and hence their parameters cannot be compared with \eqref{eq:gkt-endpoint} directly.
Very recently, the parameters of the family of impure $\dsb{3(2\delta-1),1,2\delta}_q$ qLRCs with
$(\delta,\delta)$-locality and $(2\delta-2)\mid (q-1)$ originally introduced in \cite[Proposition~6]{GHMM2026QIP}
were generalized in \cite[Proposition~32]{MS2026} to every prime power $q\ge2\delta-1$
with three disjoint $(\delta,\delta)$-recovering sets.
We easily verify that these qLRCs are GKT-optimal.
However, it is still unknown whether infinite families of GKT-optimal impure qLRCs exist
for {\em more flexible} parameters, such as $r\neq \delta$ and $\kappa>1$. 
In addition, the minimum localities of the known qLRCs for other values of $\delta$ have not been determined.
Combining the observations 2) and 3) above, we also consider the following problem.

\medskip 

\begin{problem}\label{prob:disjoint-optimal-construction}
Can we construct infinite families of GKT-optimal impure $(r,\delta)$-qLRCs
with more flexible parameters and explicit minimum localities?
\end{problem}

\end{enumerate}

\subsection{Our contributions and techniques}\label{subsec:contributions}
This paper addresses {\bf Problem}~\ref{prob:quantum-classical-locality} and {\bf Problem}~\ref{prob:disjoint-optimal-construction}
by using CSS codes constructed from generalized monomial-Cartesian codes (GMCCs)
(see Definitions~\ref{def:cartesian-code} and \ref{def:chain-gmcc} for details).
Our {\em main technique} is to group the reduced monomials of a GMCC by their degree in the last variable as follows,
where we borrow some terminology used later.
Fix integers $s\ge1$ and $\delta_0\ge2$.
We allow all monomials of degree at most $\delta_0-2$ in the last variable, and
for each of the $s$ degrees $\delta_0-1,\ldots,\delta_0+s-2$, we prescribe a GMCC in the remaining variables.
A nested pair of such GMCCs then corresponds to $s$ nested pairs of GMCCs in the remaining variables.
Both the relative minimum distance of the pair and the local recovery of the resulting CSS code can be reduced to these $s$ nested pairs.
Grouping the coordinates by the values of the remaining variables then
gives disjoint $(s+\delta_0-1,\delta_0)$-recovering sets
of the resulting CSS code.
Here, $\delta_0$ is fixed for the designed ability of quantum local recovery
with $s$ before the GMCCs in the remaining variables are prescribed.
Once we obtain the CSS code with the designed $\delta_0$,
we determine its minimum localities for {\em all admissible} $\delta\ge2$
(except for $\delta_0<\delta\le s+\delta_0-1$ when $s\ge2$)
and the quantum recovering sets that exactly attain these minimum localities,
and then decide its GKT-optimality. 

The {\em main contributions} of this paper are as follows
and some terms used here can be found in Sections~\ref{sec:preliminaries}--\ref{sec:optimality}.

\begin{enumerate}
\item In Section~\ref{sec:preliminaries}, we reduce the minimum locality of an arbitrary CSS code $\mathcal{Q}$
to local relative distances $\lambda(J)$, defined through the punctured and shortened codes on a subset $J$ of coordinate positions.
In Theorem~\ref{lem:css-support-restrictions}, we show that $J$ is an
$(r,\delta)$-recovering set of $\mathcal{Q}$ if and only if $\lambda(J)\ge\delta$,
and express $r_{\mathcal Q}^{\min}(\delta)$ through $\lambda(J)$ in Corollary~\ref{cor:locality-local-distance}. 
Focusing on $s$ nested pairs of arbitrary linear codes, we develop in Lemma~\ref{thm:coefficient-distance} 
upper and lower bounds on the relative minimum distances  
and these two bounds coincide for $s=1$ and for equal nested pairs. 

\item 
For the CSS codes \eqref{eq:gmcc-css} defined via GMCCs with decreasing defining sets,
we bound their minimum distances in Theorem~\ref{cor:gmcc-distance-specialization} based on Lemma~\ref{thm:coefficient-distance},
and present in Theorem~\ref{thm:order-comparison} an infinite family of impure $\dsb{12M,2M-4,8}_q$ CSS codes
for which our bounds give the exact minimum distance,
whereas the estimates of \cite[Theorem~16]{GGHR2018} provide a lower bound of at most $6$.
In Lemma~\ref{lem:blocks-recovering}, we prove that the blocks $J_{\mathbf x}$ \eqref{eq:disjoint-recovering-partition}
can always form disjoint $(s+\delta_0-1,\delta_0)$-recovering sets of $\mathcal{Q}$
and characterize the GKT-optimality for them.
When $\sigma=s+\delta_0$, the minimum localities for $2\le\delta\le\delta_0$ are determined in Theorem~\ref{thm:all-s1},
which also shows that under a threshold condition, every recovering set for $\delta>s+\delta_0-1$
is a union of some exact blocks based on a reduction established in Lemma \ref{lem:block-relative-distance}.

\item We further consider two classes of decreasing chains applied for CSS codes $\mathcal{Q}$ \eqref{eq:gmcc-css}
considered throughout the paper.
The first class is the chains of intervals for $m=1$ and $s\geq 1$ in \eqref{eq:profile-conditions}
and the second class takes $m\ge2$ and $s=1$ with
$A_0=\Lambda\setminus\{\boldsymbol\tau\}$ and two different choices of $B_0$ in \eqref{eq:multivariate-choice-1} and \eqref{eq:multivariate-choice-2}.
For $m=1$, the exact parameters and local relative distances of $\mathcal{Q}$ respectively follow from Lemmas~\ref{thm:corrected-family} and~\ref{lem:interval-local-distance}.
Finally, we obtain four infinite families of GKT-optimal impure qLRCs with explicit minimum localities in
Theorems~\ref{thm:optimal-attainment}, \ref{thm:two-factor-optimality}, \ref{thm:multivariate-s1}, and~\ref{thm:multivariate-s1-b},
and their conditions and parameters are listed in Table~\ref{tab:families} for easy reference.
In particular, for every admissible $\delta$ with $2\le\delta\le d$, except for $\delta_0<\delta\le s+\delta_0-1$ in Theorem~\ref{thm:two-factor-optimality},
we determine their minimum localities and the recovering sets that exactly attain them. 
We also characterize when these qLRCs have $(r,\delta)$-locality for some $r<n-\delta+1$ 
while their constituent codes have no classical $(r',\delta)$-locality for any $0<r'<n$. 
Consequently, the four families provide positive answers to {\bf Problem}~\ref{prob:disjoint-optimal-construction}.
In addition, Corollary~\ref{cor:cx-cz} presents two explicit subclasses of GKT-optimal impure qLRCs $\CSS(\C,\C)$,
which have $(r,\delta)$-locality that $\C$ does not have.
This answers {\bf Problem}~\ref{prob:quantum-classical-locality} affirmatively.

\item Comparing with the known constructions in Section~\ref{subsec:cx-cz}, we find that
taking special parameters in Theorem~\ref{thm:optimal-attainment} and settings at the beginning of Section~\ref{subsec:s-ge-2}
can immediately revisit the only known parameters of impure qLRCs in \cite{GHMM2026QIP} and \cite{MS2026}.
Therefore, for these known parameters, our construction gives impure qLRCs whose GKT-optimality and minimum localities are determined,
which were not discussed in the original papers.
Moreover, our constructions also provide parameters that the known ones do not have.
For example, Theorem~\ref{thm:two-factor-optimality} provides
GKT-optimal impure qLRCs with locality $r=s+\delta_0-1>\delta_0=\delta$
and Theorems~\ref{thm:multivariate-s1} and~\ref{thm:multivariate-s1-b} contain
GKT-optimal impure qLRCs with unbounded lengths over a fixed field. 
In particular,
these minimum localities need not be monotone in $\delta$, 
as shown in Remark~\ref{rem:nonmonotone-locality} and Table \ref{tab:nonmonotone}. 

\end{enumerate}

\begin{table}[t]
\centering
\caption{GKT-optimal impure $(r,\delta)$-qLRCs obtained in this paper}\label{tab:families}
\footnotesize
\setlength{\tabcolsep}{3pt}
\renewcommand{\arraystretch}{1.3}
\resizebox{\textwidth}{!}{%
\begin{tabular}{c|c|c|c|c|c}
\toprule
No. & $\dsb{n,\kappa,d}_q$ & $(r,\delta)$ & $r_{\mathcal Q}^{\min}(\delta)$ & Condition & Reference\\
\midrule
1 & $\dsb{h_1(2\delta_0-1),\kappa,\delta_0(\lfloor\frac{h_1-\kappa}{2}\rfloor+1)}_q$
  & $(\delta_0,\delta_0)$
  & \begin{tabular}{@{}r@{\ for\ }l@{}}$\delta_0$ & $2\le\delta\le\delta_0$,\\ $\mathcal{R}_1$  & $\delta_0<\delta\le d$\end{tabular}
  & \makecell[c]{$h_1\ge3$, $1\le\kappa\le h_1-2$,\\ $q\ge\max\{h_1,~2\delta_0-1\}$}
  & Theorem~\ref{thm:optimal-attainment}\\ \hline
2 & $\dsb{h_1(s+2\delta_0-2),s(h_1-2u)+2,u(s+\delta_0-1)}_q$
  & $(s+\delta_0-1,\delta_0)$
  & \begin{tabular}{@{}r@{\ for\ }l@{}}$s+\delta_0-1$ & $2\le\delta\le\delta_0$, \\ $\mathcal{R}_2$ & $s+\delta_0-1<\delta\le d$\end{tabular}
  & \makecell[c]{$s,~u\ge2$, $h_1\ge2u+1$, $\delta_0\ge u(s-1)$,\\ $q\ge\max\{h_1,~s+2\delta_0-2\}$}
  & Theorem~\ref{thm:two-factor-optimality}\\ \hline
3 & $\dsb{(2\delta_0-1)\prod_{j=1}^mh_j,\prod_{j=1}^mh_j-2,2\delta_0}_q$
  & $(\delta_0,\delta_0)$
  & \begin{tabular}{@{}r@{\ for\ }l@{}}$\delta_0$ & $2\le\delta\le\delta_0$,\\ $n-\delta+1$ & $\delta_0<\delta\le d$\end{tabular}
  & \makecell[c]{$m\ge2$, $h_1,\ldots,h_m\ge2$,\\ $q\ge\max\{h_1,\ldots,h_m,~2\delta_0-1\}$}
  & Theorem~\ref{thm:multivariate-s1}\\ \hline
4 & $\dsb{(2\delta_0-1)\prod_{j=1}^mh_j,\prod_{j=1}^mh_j-3,2\delta_0}_q$
  & $(\delta_0,\delta_0)$
  & \begin{tabular}{@{}r@{\ for\ }l@{}}$\delta_0$ & $2\le\delta\le\delta_0$,\\ $n-\delta+1$ & $\delta_0<\delta\le d$\end{tabular}
  & \makecell[c]{$m\ge2$, $h_1,\ldots,h_m\ge2$,\\ $q\ge\max\{h_1,\ldots,h_m,~2\delta_0-1\}$}
  & Theorem~\ref{thm:multivariate-s1-b}\\
\bottomrule
\end{tabular}}

\vspace{1mm}
\begin{minipage}{0.96\textwidth}
\footnotesize
\raggedright
\textit{Note.}
In this table,
$\mathcal{R}_1=(2\delta_0-1)(\lceil(h_1+\kappa)/2\rceil+\lceil\delta/\delta_0\rceil-1)-\delta+1$
and
$\mathcal{R}_2=(s+2\delta_0-2)(h_1-u+\lceil\delta/(s+\delta_0-1)\rceil)-\delta+1$.
\end{minipage}
\end{table}

The rest of this paper is organized as follows. Section~\ref{sec:preliminaries}
recalls quantum $(r,\delta)$-locality, the CSS construction, and GMCCs,
where we also provide a characterization of the minimum locality of an arbitrary CSS code in terms of its local relative distances.
Section~\ref{sec:multivariate} introduces the CSS codes derived from GMCCs considered throughout the paper, gives a general bound on relative minimum distances through nested pairs of linear codes,
applies it to these CSS codes, and studies their recovering sets and minimum localities.
Section~\ref{sec:optimality} answers {\bf Problem}~\ref{prob:quantum-classical-locality} and
{\bf Problem}~\ref{prob:disjoint-optimal-construction}
by constructing four infinite families of GKT-optimal impure qLRCs with explicit minimum localities,
and compares them with known constructions.
Finally, Section~\ref{sec:conclusion} concludes the paper.

\section{Preliminaries}\label{sec:preliminaries}
In this section, we first recall recovering sets,
the CSS construction of $(r,\delta)$-qLRCs, and GMCCs.
We also establish the relationship between the minimum locality of a CSS code and its local relative distances.
In particular, we only consider $\dsb{n,\kappa>0,d}_q$ CSS codes throughout the paper,
since $\dsb{n,0,d}_q$ CSS codes must be pure \cite{KKKS2006}.
We also need the following notation.

Let $\F_q^*=\F_q\setminus\{0\}$ and write $[n]=\{1,\ldots,n\}$.
Let $\mathbf{0}$ and $\mathbf{1}$ denote the {\em all-zero} and {\em all-one} vectors of appropriate lengths,
respectively.
Given any vector $\mathbf{c}=(c_1,\ldots,c_n)$ over $\F_q$, we denote
its {\em support}, {\em Hamming weight}, and {\em restriction} on the subset $J\subseteq[n]$
by $\supp(\mathbf{c})=\{i:\,c_i\neq 0\}$,
$\wt(\mathbf{c})=|\supp(\mathbf{c})|$,
and $\mathbf{c}|_J$, respectively.
An $[n,k,d]_q$ {\em linear code} $\C$ is a $k$-dimensional linear subspace of $\F_q^n$ with minimum Hamming distance
$d(\C)=\min\{\wt(\mathbf{c}):\,\mathbf{c}\in\C,~\mathbf{c}\neq\mathbf{0}\}$.
An $[n,k,n-k+1]_q$ linear code is further called a {\em maximum distance separable} (MDS) code.
A \emph{minimal support} of a linear code $\C$ is a set $\mathcal{K}=\supp(\mathbf z)$ with $\mathbf 0\ne\mathbf z\in\C$ and $\{\mathbf z'\in\C:\,\supp(\mathbf z')\subsetneq\mathcal{K}\}=\{\mathbf 0\}$.
Therefore, each support of a minimum-weight codeword is a minimal support,
but the converse is not necessarily true.

For a subcode $\D\subsetneq\C$, the \emph{relative minimum distance} of $\C$
with respect to $\D$ is
$d(\C\setminus\D)=\min\{\wt(\mathbf{c}):\,\mathbf{c}\in\C\setminus\D\}$.
In particular, we stipulate $d(\{\mathbf{0}\})=\infty$ and $d(\C\setminus\C)=\infty$,
and the minimum of an empty set is regarded as $\infty$.
Given two vectors $\mathbf{v},~\mathbf{w}\in(\F_q^*)^n$,
let $\diag(\mathbf{v})$ be the {\em diagonal matrix} with diagonal $\mathbf{v}$,
and let $\mathbf{w}/\mathbf{v}$ be their {\em componentwise quotient}.
Define $\C\diag(\mathbf{v})=\{\mathbf{c}\cdot \diag(\mathbf{v}):\,\mathbf{c}\in\C\}$,
which is {\em monomially equivalent} to $\C$.
Given a nonempty set $J\subseteq[n]$ and an $[n,k]_q$ linear code $\C$, we call
$P_J(\C)=\{\mathbf{c}|_J:\,\mathbf{c}\in\C\}$ and
$S_J(\C)=\{\mathbf{c}|_J:\,\mathbf{c}\in\C,~\supp(\mathbf{c})\subseteq J\}$
the {\em punctured code} and the {\em shortened code} of $\C$ on $J$, respectively.

\subsection{The CSS construction of qLRCs and minimum locality}\label{subsec:quantum-locality-css}

Following \cite{GHMM2026FFA,GKT2026,KKKS2006}, we recall
basic definitions of quantum $(r,\delta)$-locality and the CSS construction of $(r,\delta)$-qLRCs.

\begin{definition}\label{def:quantum-locality}{\rm (\!\! \cite[Definitions 9 and 10]{GHMM2026FFA})}
Let $r\ge1$ and $\delta\ge2$ be integers.
A quantum code $\mathcal Q\subseteq(\mathbb C^q)^{\otimes n}$ is said to be
an \emph{$(r,\delta)$-quantum locally recoverable code}, abbreviated as $(r,\delta)$-{\em qLRC},
if for each index $i\in[n]$, there exists a set $J\subseteq[n]$ containing $i$
with $\delta\le|J|\le r+\delta-1$ such that for every subset $I\subseteq J$ of cardinality $\delta-1$,
there exists a trace-preserving quantum operation $\mathcal R_{\mathcal Q,I}^{J}$, which acts only on the qudits corresponding to $J$ and keeps untouched the remaining ones, satisfying
$
\mathcal R_{\mathcal Q,I}^{J}\circ\Gamma^I(|\varphi\rangle\langle\varphi|)=|\varphi\rangle\langle\varphi|
$
for any $|\varphi\rangle\in\mathcal Q$,
where $\Gamma^I$ denotes the completely depolarizing channel on the qudits in $I\subseteq[n]$.
\end{definition}

As in \cite{GKT2026}, we call the set $J$ in Definition~\ref{def:quantum-locality}
an \emph{$(r,\delta)$-recovering set} of $\mathcal Q$.
Note also that a set $J\subseteq[n]$ with $\delta\le|J|\le r+\delta-1$ is a
{\em classical $(r,\delta)$-recovering set} of a linear code $\C$ if $d(P_J(\C))\ge\delta$.
In particular, for a nontrivial linear code $\C$,
there is some $r$ for which $\C$ has classical $(r,\delta)$-locality if and only if $\delta\le d(\C)$.
To facilitate potential parallel erasure recovery in a qLRC,
we further consider disjoint recovering sets.

\begin{definition}\label{def:recovering-sets}
Let $r\ge1$ and $\delta\ge2$ be integers. A set $J\subseteq[n]$ with $\delta\le|J|\le r+\delta-1$ is an \emph{$(r,\delta)$-recovering set} of $\mathcal Q$
if the recovery operation in Definition~\ref{def:quantum-locality} exists for every $I\subseteq J$ with $|I|=\delta-1$.
Moreover,
\begin{itemize}
  \item the set $J$ is \emph{proper} if $J\subsetneq[n]$, and
  \item the  code $\mathcal Q$ has \emph{disjoint $(r,\delta)$-recovering sets} if each coordinate $i\in[n]$ lies in an $(r,\delta)$-recovering set $J(i)$ with $|J(i)|=r+\delta-1$
such that for all $i,~j\in[n]$, either $J(i)=J(j)$ or $J(i)\cap J(j)=\varnothing$.
\end{itemize}
\end{definition}

\begin{lemma}\label{lem:css-construction}{\rm (\!\! CSS construction \cite{KKKS2006})}
Let $\C_X$ and $\C_Z$ be $[n,k_X]_q$ and $[n,k_Z]_q$ linear codes with $\C_Z^\perp\subsetneq\C_X$.
Then the CSS code $\mathcal Q=\CSS(\C_X,\C_Z)$ is an $\dsb{n,k_X+k_Z-n,d}_q$
quantum stabilizer code  with
\[
 d=\min\{d(\C_X\setminus\C_Z^\perp),~d(\C_Z\setminus\C_X^\perp)\}.
\]
\end{lemma}
For brevity, we denote $d_{XZ}=d(\C_X\setminus\C_Z^\perp)$ and $d_{ZX}=d(\C_Z\setminus\C_X^\perp)$.
Hence, the minimum distance of $\mathcal Q$ is $d=\min\{d_{XZ},~d_{ZX}\}$
and the minimum weight of a nonidentity stabilizer element of $\mathcal Q$ is
$\sigma=\min\{d(\C_X^\perp),~d(\C_Z^\perp)\}$,
since the stabilizer is generated by the operators associated with $\C_Z^\perp$ and $\C_X^\perp$ (see also \cite{KKKS2006}).
Since $\C_Z^\perp\subseteq\C_X$ and $\C_X^\perp\subseteq\C_Z$, we have $d(\C_X)=\min\{d_{XZ},~d(\C_Z^\perp)\}$ and $d(\C_Z)=\min\{d_{ZX},~d(\C_X^\perp)\}$,
and hence $\min\{d(\C_X),~d(\C_Z)\}=\min\{d,~\sigma\}$.
Therefore, the condition $\sigma<d$ for $\mathcal Q$ to be {\em impure} is equivalent to $d>\min\{d(\C_X),~d(\C_Z)\}$, as stated in Section~\ref{sec:introduction}.

Based on Lemma~\ref{lem:css-construction},
whether a CSS code $\CSS(\C_X,\C_Z)$ has $(r,\delta)$-locality can be determined
by the following relation between puncturing and shortening of constituent codes
$P_J(\C)$ and $S_J(\C)$.

\begin{lemma}\label{lem:css-recovery-criterion}{\rm (\!\! \cite[Proposition~26]{GHMM2026FFA})}
Let $\mathcal Q=\CSS(\C_X,\C_Z)$ be a CSS code, and let $r\ge1$ and $\delta\ge2$ be integers.
A set $J\subseteq[n]$ with $\delta\le|J|\le r+\delta-1$ is an $(r,\delta)$-recovering set of $\mathcal Q$
if and only if
\begin{equation}\label{eq:css-recovery-criterion}
 S_I(P_J(\C_X))=S_I(\C_Z^\perp)~~{\rm and}~~
 S_I(P_J(\C_Z))=S_I(\C_X^\perp)
\end{equation}
for every $I\subseteq J$ with $|I|=\delta-1$.
Moreover, $\mathcal Q$ has $(r,\delta)$-locality
if such recovering sets cover all coordinates in $[n]$.
\end{lemma}

Since $\C_Z^\perp\subseteq\C_X$ and $\C_X^\perp\subseteq\C_Z$, the shortened codes $S_J(\C_Z^\perp)$ and $S_J(\C_X^\perp)$ are subcodes of $P_J(\C_X)$ and $P_J(\C_Z)$, respectively.
Let $\mathcal Q=\CSS(\C_X,\C_Z)$ be an {\em arbitrary} CSS code with $\C_Z^\perp\subsetneq\C_X$.
For $\varnothing\ne J\subseteq[n]$, define the \emph{local relative distance}
\begin{equation}\label{eq:general-local-relative-distance}
 \lambda(J)=\min\{d(P_J(\C_X)\setminus S_J(\C_Z^\perp)),~
                   d(P_J(\C_Z)\setminus S_J(\C_X^\perp))\}.
\end{equation}
With Lemma~\ref{lem:css-recovery-criterion},
we derive characterizations on when a CSS code has a prescribed recovering set
using its local relative distance and minimal supports of the corresponding dual codes.

\begin{theorem}\label{lem:css-support-restrictions}
Let $\mathcal Q=\CSS(\C_X,\C_Z)$ be an $\dsb{n,\kappa,d}_q$ CSS code with $\C_Z^\perp\subsetneq\C_X$,
and let $r\ge1$ and $\delta\ge2$ be integers.
Then a set $J$ with $\delta\le|J|\le r+\delta-1$ is
an $(r,\delta)$-recovering set of $\mathcal Q$ if and only if $\lambda(J)\ge\delta$.
Moreover, the following results hold.
\begin{enumerate}[label=\arabic*),leftmargin=*,itemsep=0pt,topsep=1pt,beginpenalty=10000]
\item $\mathcal Q$ has $(r',\delta)$-locality for some $r'$ if and only if
$\delta\le d$.

\item Let $\mathcal{K}$ be a minimal support of $\C_X^\perp$ or of $\C_Z^\perp$ with $|\mathcal{K}|\le\delta$. Then every
$(r,\delta)$-recovering set $J$ of $\mathcal Q$ satisfies $\mathcal{K}\subseteq J$ or
$\mathcal{K}\cap J=\varnothing$.
\end{enumerate}
\end{theorem}
\begin{IEEEproof}
Let $J\subseteq[n]$ satisfy $\delta\le|J|\le r+\delta-1$.
By Lemma~\ref{lem:css-recovery-criterion}, $J$ is not $(r,\delta)$-recovering if and only if at least one equality in \eqref{eq:css-recovery-criterion} fails for some $I\subseteq J$ with $|I|=\delta-1$.
We first show that this occurs for a fixed $I$ if and only if some codeword in $P_J(\C_X)\setminus S_J(\C_Z^\perp)$ or in $P_J(\C_Z)\setminus S_J(\C_X^\perp)$ is supported in $I$.

Take the first equality in \eqref{eq:css-recovery-criterion} as an example and the second equality there can be treated in the same way.
Since $\C_Z^\perp\subseteq\C_X$, we have $S_I(\C_Z^\perp)\subseteq S_I(P_J(\C_X))$.
Every element of $S_I(P_J(\C_X))$ has the form $\mathbf c|_I$ for a unique $\mathbf c\in P_J(\C_X)$ supported in $I$.
For such $\mathbf c$, suppose first that $\mathbf c|_I=\mathbf z|_I$ for some $\mathbf z\in\C_Z^\perp$
with $\supp(\mathbf z)\subseteq I$.
Since $\mathbf c$ and $\mathbf z|_J$ both vanish on $J\setminus I$, we have $\mathbf c=\mathbf z|_J\in S_J(\C_Z^\perp)$.
Conversely, suppose that $\mathbf c=\mathbf z|_J$ for some $\mathbf z\in\C_Z^\perp$ supported in $J$.
Then $\supp(\mathbf z)=\supp(\mathbf c)\subseteq I$, and hence $\mathbf c|_I=\mathbf z|_I\in S_I(\C_Z^\perp)$.
Therefore, $\mathbf c|_I\in S_I(\C_Z^\perp)$ if and only if $\mathbf c\in S_J(\C_Z^\perp)$.
Together with $S_I(\C_Z^\perp)\subseteq S_I(P_J(\C_X))$, this shows that
the first equality fails if and only if some $\mathbf c\in P_J(\C_X)$
supported in $I$ does not lie in $S_J(\C_Z^\perp)$, which proves the claim.

Hence, $J$ is not $(r,\delta)$-recovering if and only if there is a vector in $P_J(\C_X)\setminus S_J(\C_Z^\perp)$ or in $P_J(\C_Z)\setminus S_J(\C_X^\perp)$ whose support is a subset of some $I\subseteq J$ with $|I|=\delta-1$.
Since $|J|\ge\delta$, this holds if and only if one of these two sets contains a vector of weight at most $\delta-1$.
Using \eqref{eq:general-local-relative-distance}, this is equivalent to $\lambda(J)\le\delta-1$.

1) Suppose that $\mathcal Q$ has $(r',\delta)$-locality for some $r'$ but $d_{XZ}<\delta$.
Since $\C_Z^\perp\subsetneq\C_X$, we can choose $\mathbf{c}\in\C_X\setminus\C_Z^\perp$ of weight $d_{XZ}$.
Since $\supp(\mathbf c)\subseteq [n]$ and all $(r',\delta)$-recovering sets cover $[n]$,
we can always find a recovering set $J$ such that $J\cap\supp(\mathbf{c})\ne\varnothing$.
Then $\mathbf{c}|_J$ is a nonzero codeword of $P_J(\C_X)$ with weight less than $\delta$.
If $\mathbf{c}|_J=\mathbf{z}|_J$ for some $\mathbf{z}\in\C_Z^\perp$ supported in $J$,
then $\mathbf{c}-\mathbf{z}\in \C_X\setminus\C_Z^\perp$ and $\wt(\mathbf{c}-\mathbf{z})<d_{XZ}$,
which is impossible.
This yields that $\mathbf{c}|_J\in P_J(\C_X)\setminus S_J(\C_Z^\perp)$.
Since $\wt(\mathbf c|_J)<\delta$, it follows from \eqref{eq:general-local-relative-distance} that $\lambda(J)<\delta$.
However, $J$ is an $(r',\delta)$-recovering set, thereby
$\lambda(J)\ge\delta$ by the main conclusion proved above, a contradiction.
Exchanging $X$ and $Z$ then gives $\delta\le \min\{d_{XZ},~d_{ZX}\}= d$.

Conversely, let $\delta\le d$. Since $P_{[n]}(\C)=S_{[n]}(\C)=\C$ for every linear code $\C$ of length $n$,
\eqref{eq:general-local-relative-distance} gives
\[
 \lambda([n])=\min\{d(\C_X\setminus\C_Z^\perp),~d(\C_Z\setminus\C_X^\perp)\}=\min\{d_{XZ},~d_{ZX}\}=d\ge\delta.
\]
By the main conclusion, the set $J=[n]$ is an $(n-\delta+1,\delta)$-recovering set.

2) Suppose that $\mathcal{K}=\supp(\mathbf z)$ is a minimal support of $\C_Z^\perp$,
and that an $(r,\delta)$-recovering set $J$ satisfies
$J\cap \mathcal{K}\ne\varnothing$ and $\mathcal{K}\not\subseteq J$. Let $I=J\cap \mathcal{K}$. Then
$1\le|I|<|\mathcal{K}|\le\delta$. Since $\C_Z^\perp\subseteq\C_X$,
$\mathbf{z}|_J$ is a nonzero codeword of $P_J(\C_X)$ with support $I$ and weight at most $\delta-1$.
If $\mathbf z|_J=\mathbf z'|_J$ for some $\mathbf z'\in\C_Z^\perp$ supported in $J$,
then $\mathbf z'$ is nonzero and $\supp(\mathbf z')=I\subsetneq\mathcal{K}$,
contradicting the minimality of $\mathcal{K}$.
Hence, $\mathbf z|_J\in P_J(\C_X)\setminus S_J(\C_Z^\perp)$
and $\lambda(J)\le\delta-1$, a contradiction again.
The analysis of the case of $\C_X^\perp$ is similar and we omit the details here.
We have completed the whole proof. 
\end{IEEEproof}
\medskip

By Theorem~\ref{lem:css-support-restrictions}, the minimum locality of a CSS code
is determined by its local relative distances as follows.
\begin{corollary}\label{cor:locality-local-distance}
Let $\mathcal Q=\CSS(\C_X,\C_Z)$ be an $\dsb{n,\kappa,d}_q$ CSS code with $\C_Z^\perp\subsetneq\C_X$
and minimum stabilizer weight $\sigma$,
and let $2\le\delta\le d$.
Then the following results hold.
\begin{enumerate}[label=\arabic*),leftmargin=*,itemsep=0pt,topsep=1pt,beginpenalty=10000]
\item The minimum locality of $\mathcal Q$ is
\[
 r_{\mathcal Q}^{\min}(\delta)=\max\bigl\{\min\{|J|:\,i\in J\subseteq[n],~|J|\ge\delta,~\lambda(J)\ge\delta\}:\,i\in[n]\bigr\}-\delta+1.
\]
\item If $\delta\le\sigma$, then for every integer $r\ge1$, the quantum $(r,\delta)$-recovering sets
are exactly the classical $(r,\delta)$-recovering sets of both $\C_X$ and $\C_Z$.
In particular,
$$r_{\mathcal Q}^{\min}(\delta)\ge\max\{r_{\C_X}^{\min}(\delta),~r_{\C_Z}^{\min}(\delta)\},$$
with equality if $\C_Z\subseteq\C_X\diag(\mathbf v)$ or $\C_X\subseteq\C_Z\diag(\mathbf v)$
for some $\mathbf v\in(\F_q^*)^n$.
\end{enumerate}
\end{corollary}
\begin{IEEEproof}
1) For each $i\in[n]$, let
$
 \ell_i=\min\{|J|:\,i\in J\subseteq[n],~|J|\ge\delta,~\lambda(J)\ge\delta\}.
$
Since $\lambda([n])=d\ge\delta$, the set $[n]$ belongs to the set in the definition of $\ell_i$.
Hence, $\ell_i$ is well-defined and $\delta\le\ell_i\le n$.
Let 
$$r_{\max}=\max\{\ell_i:\,i\in[n]\}-\delta+1.$$ 
Then $r_{\max}\ge1$.
By Theorem~\ref{lem:css-support-restrictions}, a set $J$ containing a coordinate $i$
is an $(r,\delta)$-recovering set if and only if $\delta\le|J|\le r+\delta-1$ and $\lambda(J)\ge\delta$.

For any given $i\in[n]$, since the set in the definition of $\ell_i$ is finite and nonempty,
we can choose a set $J_i$ attaining $\ell_i$,
that is, $i\in J_i$, $|J_i|=\ell_i$ and $\lambda(J_i)\ge\delta$.
Since $\delta\le|J_i|=\ell_i\le r_{\max}+\delta-1$,
the set $J_i$ is an $(r_{\max},\delta)$-recovering set containing $i$,
implying that $r_{\mathcal Q}^{\min}(\delta)\le r_{\max}$.
Conversely, let $r$ be an integer with $1\le r<r_{\max}$.
By the definition of $r_{\max}$, there exists $i^*\in[n]$ such that $\ell_{i^*}=r_{\max}+\delta-1$.
Note that every set $J$ with $i^*\in J$, $|J|\ge\delta$ and $\lambda(J)\ge\delta$
satisfies $|J|\ge\ell_{i^*}>r+\delta-1$.
Therefore, no $(r,\delta)$-recovering set contains $i^*$.
It follows that $\mathcal Q$ does not have $(r,\delta)$-locality and
$r_{\mathcal Q}^{\min}(\delta)\ge r_{\max}$.
Combining these two aspects, we conclude that $r_{\mathcal Q}^{\min}(\delta)=r_{\max}$.

2) Let $\delta\le\sigma$, and let $J\subseteq[n]$.
Every nonzero codeword of $S_J(\C_Z^\perp)$ or of $S_J(\C_X^\perp)$ has weight at least
$\sigma\ge\delta$.
Hence, $\lambda(J)\ge\delta$ if and only if $d(P_J(\C_X))\ge\delta$ and $d(P_J(\C_Z))\ge\delta$.
By Theorem~\ref{lem:css-support-restrictions}, this gives the first assertion
and further implies that $r_{\mathcal Q}^{\min}(\delta)\ge\max\{r_{\C_X}^{\min}(\delta),~r_{\C_Z}^{\min}(\delta)\}.$

If $\C_Z\subseteq\C_X\diag(\mathbf v)$, then $P_J(\C_Z)\subseteq P_J(\C_X)\diag(\mathbf v|_J)$,
and hence $d(P_J(\C_Z))\ge d(P_J(\C_X))$ for every $J$.
Then every classical $(r,\delta)$-recovering set of $\C_X$ is a classical $(r,\delta)$-recovering set of $\C_Z$,
and the quantum $(r,\delta)$-recovering sets are exactly the classical $(r,\delta)$-recovering sets of $\C_X$.
It follows then that $r_{\mathcal Q}^{\min}(\delta)=r_{\C_X}^{\min}(\delta)\ge r_{\C_Z}^{\min}(\delta)$.
The case $\C_X\subseteq\C_Z\diag(\mathbf v)$ is symmetric.
\end{IEEEproof}

\subsection{Basic properties of GMCCs with decreasing sets}\label{subsec:gmcc-general}

Let $\mathcal P=\prod_{j=1}^t\mathcal S_j\subseteq\F_q^t$ be the {\em Cartesian product} of
$t$ nonempty sets $\mathcal S_j\subseteq\F_q$, called the \emph{evaluation sets}, with $|\mathcal S_j|=h_j$ for $1\le j\le t$.
Fix an enumeration $\mathcal P=\{\mathbf{p}_1,\ldots,\mathbf{p}_n\}$
with $n=\prod_{j=1}^t h_j$.
For $\mathcal H=\prod_{j=1}^t\{0,\ldots,h_j-1\}$ and $\mathbf{i}=(i_1,\ldots,i_t)\in \mathcal H$,
we write $\mathbf{X}^{\mathbf{i}}=X_1^{i_1}\cdots X_t^{i_t}$.
For $\Omega\subseteq \mathcal H$, let
$\mathcal L(\Omega)=\Span_{\F_q}\{\mathbf{X}^{\mathbf{i}}:\,\mathbf{i}\in\Omega\}$
be the {\em linear space} of polynomials with monomial exponents in $\Omega$.
For a polynomial $f\in \mathcal L(\Omega)$,
its {\em evaluation vector} with respect to the enumeration of $\mathcal P$ is
$\ev_{\mathcal P}(f)=(f(\mathbf{p}_1),\ldots,f(\mathbf{p}_n))$,
abbreviated to $\ev(f)$ when $\mathcal P$ is clear from the context.
We call $f$ \emph{reduced} if $\deg_{X_j}(f)<h_j$ for every $j$.

\begin{definition}\label{def:cartesian-code}
For a set $\Omega\subseteq \mathcal H$ and $\mathbf{v}=(v_1,\ldots,v_n)\in(\F_q^*)^n$,
the \emph{generalized monomial-Cartesian code (GMCC)} with \emph{defining set} $\Omega$
and \emph{multiplier} $\mathbf{v}$ is defined by
\[
 \C_{\mathcal P}(\Omega,\mathbf{v})=\bigl\{(v_1f(\mathbf{p}_1),\ldots,v_nf(\mathbf{p}_n)):\,f\in\mathcal L(\Omega)\bigr\}\subseteq\F_q^n.
\]
For $\mathbf{v}=\mathbf{1}$, it is further called a \emph{monomial-Cartesian code}
and is abbreviated as
$\C_{\mathcal P}(\Omega)=\C_{\mathcal P}(\Omega,\mathbf{1})$.
In addition, $\C_{\mathcal P}(\Omega,\mathbf{v})=\C_{\mathcal P}(\Omega)\diag(\mathbf{v})$
and $\C_{\mathcal P}(\varnothing,\mathbf{v})=\{\mathbf{0}\}$.
\end{definition}

To describe Euclidean duals of GMCCs, we need the following notation.
For a nonempty set $\mathcal S\subseteq\F_q$, put
$f_{\mathcal S}(Z)=\prod_{z\in\mathcal S}(Z-z)$.
For $\mathbf{p}_i\in\mathcal P$ with $1\leq i\leq n$ and any $\boldsymbol\xi=(\xi_1,\ldots,\xi_t)\in\mathcal P$,
define
\begin{align}\label{eq.w}
 \mathbf{w}=(w(\mathbf{p}_1),\ldots,w(\mathbf{p}_n)),~
{\rm where}~
w(\boldsymbol\xi)=\prod_{j=1}^t\prod_{z\in\mathcal S_j\setminus\{\xi_j\}}(\xi_j-z)^{-1}
 =\prod_{j=1}^t f'_{\mathcal S_j}(\xi_j)^{-1}.
\end{align}
Recall that the length and dimension of a monomial-Cartesian code were determined in \cite[Proposition~2.1]{LMS2020},
and its minimum distance and its dual for a decreasing set $\Omega$ in \cite[Theorems~3.9 and~3.3]{CLMS2021}.
Here, a set $\Omega\subseteq\mathcal H$ is \emph{decreasing} if $\mathbf{i}\in\Omega$ implies $\mathbf{j}\in\Omega$
for every $\mathbf{0}\le\mathbf{j}\le\mathbf{i}$ coordinatewise.
Since $\C_{\mathcal P}(\Omega,\mathbf{v})$ is monomially equivalent to $\C_{\mathcal P}(\Omega)$
and $(\C\diag(\mathbf{v}))^\perp=\C^\perp\diag(\mathbf{v})^{-1}$ for every linear code $\C$,
all these results give the following lemma for GMCCs with decreasing defining sets.

\begin{lemma}\label{lem:cartesian-duality-distance}
For a nonempty decreasing set $\Omega\subseteq \mathcal H$ and a multiplier $\mathbf{v}\in(\F_q^*)^n$,
the following results hold.
\begin{enumerate}[label=\arabic*),leftmargin=*,itemsep=0pt,topsep=1pt,beginpenalty=10000]
\item The GMCC $\C_{\mathcal P}(\Omega,\mathbf{v})$ has parameters
$
\left[\prod_{j=1}^t h_j,|\Omega|,\min\Bigl\{\prod_{j=1}^t(h_j-i_j):\,\mathbf{i}=(i_1,\ldots,i_t)\in\Omega\Bigr\}\right]_q.
$
\item $\C_{\mathcal P}(\Omega,\mathbf{v})^\perp=\C_{\mathcal P}(\Omega^\sharp,\mathbf{w}/\mathbf{v}),$
where $\Omega^\sharp=\bigl\{(h_1-1-i_1,\ldots,h_t-1-i_t):\,\mathbf{i}=(i_1,\ldots,i_t)\in \mathcal H\setminus\Omega\bigr\}.$
Moreover, $\C_{\mathcal P}(\Omega)^\perp=\C_{\mathcal P}(\Omega^\sharp)\,\diag(\mathbf{w})$.
\end{enumerate}
\end{lemma}

Recall that a \emph{monomial order} $\prec$ is a total order on the monomials of $\F_q[X_1,\ldots,X_t]$
such that $1\preceq\mathbf X^{\mathbf i}$ for every monomial $\mathbf X^{\mathbf i}$,
and $\mathbf X^{\mathbf i}\prec\mathbf X^{\mathbf j}$ implies
$\mathbf X^{\mathbf i}\mathbf X^{\mathbf k}\prec\mathbf X^{\mathbf j}\mathbf X^{\mathbf k}$
for every monomial $\mathbf X^{\mathbf k}$.
We compare exponents $\mathbf i,~\mathbf j\in \mathcal H$ through their monomials,
writing $\mathbf i\prec\mathbf j$ if $\mathbf X^{\mathbf i}\prec\mathbf X^{\mathbf j}$,
and $\min_\prec$ and $\max_\prec$ denote the least and the largest element in this order.
For two nested GMCCs, two direct estimates of the relative minimum distance with respect to a monomial order
are available in the literature.

\begin{lemma}\label{lem:cartesian-relative-estimates}{\rm (\!\! \cite[Theorem~16]{GGHR2018})}
Let $\prec$ be a monomial order and let $\Omega_2\subsetneq\Omega_1\subseteq \mathcal H$ be nested defining sets. Then
\[
 \begin{aligned}
 d(\C_{\mathcal P}(\Omega_1)\setminus\C_{\mathcal P}(\Omega_2))&\ge\min\Bigl\{\prod_{j=1}^t(h_j-i_j):\, \mathbf i\in\Omega_1,~\mathbf i\succeq\min\nolimits_\prec(\Omega_1\setminus\Omega_2)\Bigr\},~{\rm and}\\
 d(\C_{\mathcal P}(\Omega_2)^\perp\setminus\C_{\mathcal P}(\Omega_1)^\perp)&\ge\min\Bigl\{\prod_{j=1}^t(i_j+1):\, \mathbf i\in \mathcal H\setminus\Omega_2,~\mathbf i\preceq\max\nolimits_\prec(\Omega_1)\Bigr\}.
 \end{aligned}
\]
\end{lemma}

\section{Minimum distances and minimum localities of CSS codes from GMCCs}\label{sec:multivariate}
In this section, we construct CSS codes from GMCCs and study their minimum distances and minimum localities.
By Lemma~\ref{lem:css-construction} and Corollary~\ref{cor:locality-local-distance},
these two quantities are determined by the relative minimum distances $d_{XZ}$ and $d_{ZX}$
and by the local relative distances $\lambda(J)$, respectively.
In Section~\ref{subsec:gmcc-css}, we introduce the GMCCs with nested decreasing chains
considered throughout this paper and the resulting CSS codes.
In Section~\ref{subsec:gmcc-distance}, we first bound the relative minimum distances of nested pairs of codes built from general linear codes,
and then apply this bound to GMCCs and compare the new bounds with known ones in the literature.
In Section~\ref{subsec:gmcc-locality}, we characterize the recovering sets of such specific CSS codes
and determine their minimum localities.

\subsection{CSS codes from GMCCs with nested decreasing chains}\label{subsec:gmcc-css}
According to Lemma~\ref{lem:css-construction}, the CSS construction requires the dual-containing condition
$\C_Z^\perp\subseteq\C_X$.
However, two GMCCs are not nested in general.
As stated in Lemma~\ref{lem:cartesian-duality-distance}.2),
the dual of a GMCC is again a GMCC when the defining set is decreasing,
and hence the required nested condition reduces to a containment of defining sets.
More precisely, if $\C_X=\C_{\mathcal P}(\Omega_1)$ and
$\C_Z=\C_{\mathcal P}(\Omega_2)\diag(\mathbf w)$ with decreasing $\Omega_1,~\Omega_2$, then
$\C_Z^\perp=\C_{\mathcal P}(\Omega_2^\sharp)$, and
$\C_Z^\perp\subseteq\C_X$ if and only if $\Omega_2^\sharp\subseteq\Omega_1$.
Since $\Omega_1^\sharp$ and $\Omega_2^\sharp$ are again decreasing,
Lemma~\ref{lem:cartesian-duality-distance}.1) also gives the minimum distances of
$\C_X$, $\C_Z$, and their duals,
which is the key to deciding whether the resulting CSS code is impure
and is helpful in determining the local relative distances and minimum localities of these CSS codes.

As a result, the CSS codes in this paper are constructed from the following GMCCs, in which the last variable is distinguished.
Fix integers $m\ge1$, $s\ge1$, and $\delta_0\ge2$. We use the notation of Section~\ref{subsec:gmcc-general}
with $t=m+1$, $h_j\ge2$ for $1\le j\le m$, and $h_{m+1}=s+2(\delta_0-1)$.
Here $\delta_0$ is fixed for the designed locally recoverable ability,
whereas Section~\ref{subsec:gmcc-locality} studies the minimum localities of the resulting CSS codes for all admissible $\delta\ge2$.
Write $\mathbf X=(X_1,\ldots,X_m)$ and let
\begin{equation}\label{eq:gmcc-notation}
\begin{gathered}
 \mathcal X=\prod_{j=1}^m\mathcal S_j,~~\Lambda=\prod_{j=1}^m\{0,\ldots,h_j-1\},~~\mathbf X^{\boldsymbol\alpha}=X_1^{\alpha_1}\cdots X_m^{\alpha_m},~~{\rm and}\\
 \rho(\boldsymbol\alpha)=\boldsymbol\tau-\boldsymbol\alpha~~{\rm for}~~
 \boldsymbol\tau=(h_1-1,\ldots,h_m-1)~~{\rm and}~~\boldsymbol\alpha=(\alpha_1,\ldots,\alpha_m)\in\Lambda.
\end{gathered}
\end{equation}
Then $\mathcal P=\mathcal X\times\mathcal S_{m+1}$, $n=(s+2\delta_0-2)\prod_{j=1}^mh_j$,
$\mathcal H=\Lambda\times\{0,\ldots,s+2\delta_0-3\}$, and
$\mathbf X^{(\boldsymbol\alpha,\ell)}=\mathbf X^{\boldsymbol\alpha}X_{m+1}^{\ell}$ for
$(\boldsymbol\alpha,\ell)\in\mathcal H$.
Moreover, we let $\rho(\Omega)=\{\rho(\boldsymbol\alpha):\,\boldsymbol\alpha\in\Omega\}$ for any $\Omega\subseteq\Lambda$,
and for a decreasing set $\Omega\subseteq\Lambda$, we denote by $\C_{\mathcal X}(\Omega)$ the 
monomial-Cartesian code on $\mathcal X$ with defining set $\Omega$.
Through the enumeration of $\mathcal P$, we index the coordinates of codes of length $n$ by the points of $\mathcal P$, and similarly the coordinates of the codes on $\mathcal X$ by the points of $\mathcal X$. For a vector $\mathbf c$ indexed in this way, we write $c(\mathbf x)$ for its \emph{coordinate} at the point $\mathbf x$.

Note that every decreasing set $\Omega\subseteq\mathcal H$ has the form
$\bigcup_{\ell=0}^{s+2\delta_0-3}\Omega_\ell\times\{\ell\}$ for decreasing sets
$\Omega_0\supseteq\cdots\supseteq\Omega_{s+2\delta_0-3}$ in $\Lambda$.
We take $\Omega_\ell=\Lambda$ for $\ell\le\delta_0-2$ and
$\Omega_\ell=\varnothing$ for $\ell\ge s+\delta_0-1$.
Then only the $s$ sets $\Omega_\ell$ with $\delta_0-1\le\ell\le s+\delta_0-2$ are free.
We call a sequence $\mathbf A=(A_0\supseteq\cdots\supseteq A_{s-1})$, with empty sets allowed,
of decreasing subsets of $\Lambda$, a \emph{nested decreasing chain} in $\Lambda$.
For such a chain, since the sets $A_i$ are decreasing and nested,
taking $\Omega_{\delta_0-1+i}=A_i$ for $0\le i<s$ gives the decreasing set
\[
 \Omega(\mathbf{A})=\Lambda\times\{0,\ldots,\delta_0-2\}\cup\bigcup_{i=0}^{s-1}\bigl(A_i\times\{\delta_0-1+i\}\bigr),
\]
and we call $\C(\mathbf A)=\C_{\mathcal P}(\Omega(\mathbf A))$ the \emph{GMCC with the nested decreasing chain} $\mathbf A$.

\begin{definition}\label{def:chain-gmcc}{\rm (CSS codes considered in this paper)}
Let $\mathbf A$ and $\mathbf B$ be nested decreasing chains in $\Lambda$.
Let
\begin{equation}\label{eq:gmcc-css}
 \C_X=\C(\mathbf A),~~\C_Z=\C(\mathbf B)\diag(\mathbf w),~~{\rm and}~~\mathcal Q=\CSS(\C_X,\C_Z)~~{\rm if}~~\C_Z^\perp\subsetneq\C_X,
\end{equation}
where $\mathbf w$ is the vector in \eqref{eq.w}.
This gives the family of CSS codes $\mathcal Q$ from GMCCs with nested decreasing chains $\mathbf A$ and $\mathbf B$
considered in this paper.
\end{definition}

The choice in Definition~\ref{def:chain-gmcc} has the following advantages,
which are confirmed later in Lemmas~\ref{lem:gmcc-pair} and~\ref{lem:blocks-recovering} as well as Theorem~\ref{thm:all-s1}.
\begin{itemize}
  \item The duals of these GMCCs have the same form, which makes the dual-containing condition required for
  the CSS construction only involve the free sets $A_i$ and $B_i$.

  \item It upper bounds the minimum stabilizer weight of $\mathcal{Q}$ by $s+\delta_0$.
  Hence, the CSS codes are impure whenever their minimum distance exceeds $s+\delta_0$.

  \item Such a choice also guarantees that the sets $\{\mathbf x\}\times\mathcal S_{m+1}$ with $\mathbf x\in\mathcal X$
form disjoint $(s+\delta_0-1,\delta_0)$-recovering sets,
which is helpful in determining the minimum localities and GKT-optimality of these CSS codes.
\end{itemize}

For $0\le i<s$, we also abbreviate
\begin{equation}\label{eq:dual-chains}
 D_i^A=\rho(\Lambda\setminus A_{s-1-i}),~~
 D_i^B=\rho(\Lambda\setminus B_{s-1-i}),~~{\rm and}~~
 T_i=A_i\setminus D_i^B,
\end{equation}
where the sets $D_i^A$ and $D_i^B$ form nested decreasing chains $\mathbf D^A$ and $\mathbf D^B$ in $\Lambda$, and $T_i=A_i\cap\rho(B_{s-1-i})$.
The following lemma gives the duals of $\C_X$ and $\C_Z$ and the parameters of the resulting CSS code.
It specializes Lemmas \ref{lem:css-construction} and \ref{lem:cartesian-duality-distance} to the defining sets $\Omega(\mathbf A)$ and $\Omega(\mathbf B)$.

\begin{lemma}\label{lem:gmcc-pair}
For the codes $\C_X$ and $\C_Z$ in \eqref{eq:gmcc-css}, the following results hold.
\begin{enumerate}[label=\arabic*),leftmargin=*,itemsep=0pt,topsep=1pt,beginpenalty=10000]
\item $\C_X^\perp=\C(\mathbf{D}^A)\diag(\mathbf{w})$ and $\C_Z^\perp=\C(\mathbf{D}^B)$.
\item $\C_Z^\perp\subseteq\C_X$ if and only if $D_i^B\subseteq A_i$ for $0\le i<s$,
in which case $\C_Z^\perp\ne\C_X$ if and only if $T_i\ne\varnothing$ for some $0\le i<s$.
\item If $\C_Z^\perp\subsetneq\C_X$, then $\mathcal Q$ has parameters $\dsb{(s+2\delta_0-2)\prod_{j=1}^mh_j,\kappa,d}_q$
and minimum stabilizer weight $\sigma$, where
\begin{equation}\label{eq:multivariate-logical-dimension}
 \kappa=\sum_{i=0}^{s-1}|T_i|=\sum_{i=0}^{s-1}(|A_i|+|B_i|)-s\prod_{j=1}^mh_j,~{\rm and}
\end{equation}
\begin{equation}\label{eq:multivariate-stabilizer}
 \sigma=\min\biggl\{s+\delta_0,~\min\Bigl\{(s+\delta_0-1-i)\prod_{j=1}^m(h_j-\alpha_j):\,0\le i<s,~\boldsymbol\alpha\in D_i^A\cup D_i^B\Bigr\}\biggr\}.
\end{equation}
\end{enumerate}
\end{lemma}
\begin{IEEEproof}
1) With Lemma~\ref{lem:cartesian-duality-distance}.2),
we first write $\Omega(\mathbf{A})^\sharp=\{(\rho(\boldsymbol\alpha),s+2\delta_0-3-\ell):\,(\boldsymbol\alpha,\ell)\in\mathcal H\setminus\Omega(\mathbf{A})\}$.
Since $\Lambda\times\{0,\ldots,\delta_0-2\}\subseteq\Omega(\mathbf{A})$ and $\ell\le s+\delta_0-2$ for $(\boldsymbol\alpha,\ell)\in\Omega(\mathbf{A})$, the set $\mathcal H\setminus\Omega(\mathbf{A})$ is the disjoint union of $\Lambda\times\{s+\delta_0-1,\ldots,s+2\delta_0-3\}$ and $(\Lambda\setminus A_i)\times\{\delta_0-1+i\}$ for $0\le i<s$.
As $(s+2\delta_0-3)-(s+\delta_0-1)=\delta_0-2$ and $(s+2\delta_0-3)-(\delta_0-1+i)=\delta_0-1+(s-1-i)$, the reflection map $(\boldsymbol\alpha,\ell)\mapsto(\rho(\boldsymbol\alpha),s+2\delta_0-3-\ell)$ maps these sets onto $\Lambda\times\{0,\ldots,\delta_0-2\}$ and $D^A_{s-1-i}\times\{\delta_0-1+(s-1-i)\}$, respectively.
Replacing $i$ by $s-1-i$, we obtain
\[
 \Omega(\mathbf{A})^\sharp=\Lambda\times\{0,\ldots,\delta_0-2\}\cup\bigcup_{i=0}^{s-1}\bigl(D^A_i\times\{\delta_0-1+i\}\bigr)=\Omega(\mathbf{D}^A).
\]
Similarly, we have $\Omega(\mathbf{B})^\sharp=\Omega(\mathbf{D}^B)$.
Hence, the two described duals follow from Lemma~\ref{lem:cartesian-duality-distance}.2)
by taking $\mathbf{v}=\mathbf{1}$ and $\mathbf{v}=\mathbf{w}$, respectively.

2) Since the evaluation vectors $\ev_{\mathcal P}(\mathbf X^{\boldsymbol\alpha}X_{m+1}^\ell)$
with $(\boldsymbol\alpha,\ell)\in\mathcal H$ form a basis of $\F_q^{n}$,
the inclusion $\C_Z^\perp=\C(\mathbf{D}^B)\subseteq\C(\mathbf{A})$
is equivalent to $D_i^B\subseteq A_i$ for $0\le i<s$.
Note that
$
 \left\{\ev_{\mathcal P}\!\left(\mathbf{X}^{\boldsymbol\alpha}X_{m+1}^{\delta_0-1+i}\right)
 +\C_Z^\perp:\,0\le i<s,~\boldsymbol\alpha\in T_i\right\}
$
forms a basis of $\C_X/\C_Z^\perp$.
Then $\C_Z^\perp\ne\C_X$ if and only if
some $T_i=A_i\setminus D_i^B$ is nonempty for $0\leq i<s$.

3) Since $D_i^B\subseteq A_i$ and $|D_i^B|=|\rho(\Lambda\setminus B_{s-1-i})|=|\Lambda|-|B_{s-1-i}|$,
we have $|T_i|=|A_i|-|D_i^B|=|A_i|+|B_{s-1-i}|-|\Lambda|$ for $0\le i<s$.
From $|\Omega(\mathbf A)|=|\Lambda|(\delta_0-1)+\sum_{i=0}^{s-1}|A_i|$, $|\Omega(\mathbf B)|=|\Lambda|(\delta_0-1)+\sum_{i=0}^{s-1}|B_i|$,
and $n=(s+2\delta_0-2)|\Lambda|$, Lemma~\ref{lem:css-construction} gives
\[\kappa=|\Omega(\mathbf{A})|+|\Omega(\mathbf{B})|-n
=\sum_{i=0}^{s-1}(|A_i|+|B_i|)-s|\Lambda|=\sum_{i=0}^{s-1}|T_i|.\]
Since multiplication by $\diag(\mathbf{w})$ preserves Hamming weights,
the minimum nonidentity stabilizer weight of the CSS code $\mathcal Q$ is
  $\sigma=\min\{d(\C(\mathbf{D}^A)),~d(\C(\mathbf{D}^B))\}$.
Applying Lemma~\ref{lem:cartesian-duality-distance}.1), we have
$
 d(\C(\mathbf{D}^A))=\min\Bigl\{(s+2\delta_0-2-\ell)\prod_{j=1}^m(h_j-\alpha_j):\,(\boldsymbol\alpha,\ell)\in\Omega(\mathbf{D}^A)\Bigr\}.
$
For $\ell\le\delta_0-2$, $(s+2\delta_0-2-\ell)\prod_{j=1}^m(h_j-\alpha_j)\geq s+\delta_0$
and the equality holds at $(\boldsymbol\alpha,\ell)=(\boldsymbol\tau,\delta_0-2)$.
For $\ell=\delta_0-1+i$, we have $s+2\delta_0-2-\ell=s+\delta_0-1-i$ and $\boldsymbol\alpha\in D_i^A$.
We then arrive at
\[
 d(\C(\mathbf{D}^A))=\min\biggl\{s+\delta_0,~\min\Bigl\{(s+\delta_0-1-i)\prod_{j=1}^m(h_j-\alpha_j):\,0\le i<s,~\boldsymbol\alpha\in D_i^A\Bigr\}\biggr\}.
\]
Note also that the same argument applies to $\C(\mathbf{D}^B)$ with $D_i^B$ in place of $D_i^A$.
Substituting these two formulas into $\sigma$ then immediately yields \eqref{eq:multivariate-stabilizer}.
\end{IEEEproof}

\subsection{Bounds on relative minimum distances of $\mathcal{Q}$}\label{subsec:gmcc-distance}
For a nonempty set $U\subseteq\mathcal X$ and arbitrary linear codes $F_i\subseteq E_i\subseteq\F_q^{|U|}$
with $0\le i<s$ and coordinates indexed by $U$, write $\mathbf E=(E_i)_{i=0}^{s-1}$ and $\mathbf F=(F_i)_{i=0}^{s-1}$.
Given coefficient vectors $\mathbf g_\ell=(g_\ell(\mathbf x))_{\mathbf x\in U}\in\F_q^{|U|}$ and
$\mathbf u_i=(u_i(\mathbf x))_{\mathbf x\in U}\in E_i$, let
\begin{equation}\label{eq:linear-coefficient-evaluation}
 f_{\mathbf x}(X_{m+1})=\sum_{\ell=0}^{\delta_0-2}g_\ell(\mathbf x)X_{m+1}^\ell+
        \sum_{i=0}^{s-1}u_i(\mathbf x)X_{m+1}^{\delta_0-1+i}~~{\rm for}~~\mathbf x\in U,
\end{equation}
and let $\mathscr C_{\mathcal S_{m+1}}(\mathbf E)$ be the set of all vectors $(f_{\mathbf x}(y))_{(\mathbf x,y)\in U\times\mathcal S_{m+1}}$.
We call the vectors $\mathbf g_\ell$ the \emph{lower coefficients} and the vector $\mathbf u_i$ the \emph{$i$th restricted coefficient}.
Since $\deg_{X_{m+1}}(f_{\mathbf x})\le s+\delta_0-2<s+2\delta_0-2$, each vector in $\mathscr C_{\mathcal S_{m+1}}(\mathbf E)$ determines its coefficient vectors uniquely.
These vectors depend linearly on the coefficient vectors,
and hence $\mathscr C_{\mathcal S_{m+1}}(\mathbf E)$ is an
$[(s+2\delta_0-2)|U|,\,|U|(\delta_0-1)+\sum_{i=0}^{s-1}\dim(E_i)]_q$ linear code.
Replacing $E_i$ by $F_i$ for $0\le i<s$ gives the subcode $\mathscr C_{\mathcal S_{m+1}}(\mathbf F)\subseteq\mathscr C_{\mathcal S_{m+1}}(\mathbf E)$.
The following lemma bounds the relative minimum distance
$d(\mathscr C_{\mathcal S_{m+1}}(\mathbf E)\setminus\mathscr C_{\mathcal S_{m+1}}(\mathbf F))$
in terms of the nested pairs $(E_i,F_i)$ of general linear codes.

\begin{lemma}\label{thm:coefficient-distance}
Suppose that $E_0\supseteq\cdots\supseteq E_{s-1}$ with $F_i\subsetneq E_i$ for some $0\le i<s$.
Let $\eta_i=d(E_i\setminus F_i)$ for $0\le i<s$. Then
\begin{equation}\label{eq:coefficient-distance-bound}
\begin{aligned}
  \min\Bigl\{(s+\delta_0-1-j)\max\bigl\{d(E_j),~ & \min\{\eta_k:\,k\le j\}\bigr\}:\,0\le j<s\Bigr\}\\
  &\mathrel{\le} d(\mathscr C_{\mathcal S_{m+1}}(\mathbf{E})\setminus\mathscr C_{\mathcal S_{m+1}}(\mathbf{F}))
  \le\min\{(s+\delta_0-1-i)\eta_i:\,0\le i<s\}.
 \end{aligned}
\end{equation}
Moreover, the following results hold.
\begin{enumerate}[label=\arabic*),leftmargin=*,itemsep=0pt,topsep=1pt,beginpenalty=10000]
\item The lower bound in \eqref{eq:coefficient-distance-bound} is at least
\begin{equation}\label{eq.add.1}
 \max\bigl\{\min\{(s+\delta_0-1-i)d(E_i):\,0\le i<s\},~\delta_0\min\{\eta_i:\,0\le i<s\}\bigr\}.
\end{equation}
\item If $s=1$, or if $E_i=E_0$ and $F_i=F_0$ for each $0\le i<s$, then  $d(\mathscr C_{\mathcal S_{m+1}}(\mathbf{E})\setminus\mathscr C_{\mathcal S_{m+1}}(\mathbf{F}))=\delta_0\eta_0$.
\item If $\eta_i=d(E_i)$ for each $0\le i<s$, then $d(\mathscr C_{\mathcal S_{m+1}}(\mathbf{E})\setminus\mathscr C_{\mathcal S_{m+1}}(\mathbf{F}))=\min\{(s+\delta_0-1-i)\eta_i:\,0\le i<s\}$.
\end{enumerate}
\end{lemma}
\begin{IEEEproof}
Let $\mathbf{c}=(f_{\mathbf x}(y))_{(\mathbf x,y)\in U\times\mathcal S_{m+1}}$ be an arbitrary vector in $\mathscr C_{\mathcal S_{m+1}}(\mathbf{E})
\setminus\mathscr C_{\mathcal S_{m+1}}(\mathbf{F})$, where the polynomials $f_{\mathbf x}$ are given by \eqref{eq:linear-coefficient-evaluation}.
Since the coefficients of $\mathbf{c}$ are unique and the $\mathbf{g}_\ell$ are unrestricted,
we have $\mathbf u_i\notin F_i$ for some $i$. Let $j$ be the
largest index with $\mathbf u_j\ne\mathbf{0}$. For every $\mathbf x\in\bigcup_{\nu=0}^j\supp(\mathbf u_\nu)$, the polynomial $f_{\mathbf x}$ is nonzero of degree at most $\delta_0-1+j$, and hence it has at least $s+\delta_0-1-j$ nonzero evaluations on $\mathcal S_{m+1}$.
Since every $i$ with $\mathbf u_i\notin F_i$ satisfies $i\le j$, we have $\bigl|\bigcup_{\nu=0}^j\supp(\mathbf u_\nu)\bigr|\ge\wt(\mathbf u_j)\ge d(E_j)$ and $\bigl|\bigcup_{\nu=0}^j\supp(\mathbf u_\nu)\bigr|\ge\wt(\mathbf u_i)\ge\eta_i\ge\min\{\eta_k:\,k\le j\}$ for such an $i$.
This yields that
$$\wt(\mathbf{c})\ge(s+\delta_0-1-j)\bigl|\bigcup_{\nu=0}^j\supp(\mathbf u_\nu)\bigr|\ge(s+\delta_0-1-j)\max\bigl\{d(E_j),~\min\{\eta_k:\,k\le j\}\bigr\}.$$
By the arbitrariness of $\mathbf{c}$, we obtain the {\em lower bound} in \eqref{eq:coefficient-distance-bound}.

For the {\em upper bound}, we fix $i$ with $\eta_i<\infty$.
Then $E_i\setminus F_i\neq \varnothing$ and there exists a vector
$\mathbf u=(u(\mathbf x))_{\mathbf x\in U}\in E_i\setminus F_i$ of weight $\eta_i$.
Since $\delta_0-1+i<s+2\delta_0-2=|\mathcal S_{m+1}|$, we can also choose
a subset $R\subseteq\mathcal S_{m+1}$ of size $\delta_0-1+i$.
Write
$p(X_{m+1})=\prod_{y\in R}(X_{m+1}-y)=\sum_{k=0}^{\delta_0-1+i}p_kX_{m+1}^k.$
Since $\mathbf u\in E_i\subseteq E_{i'}$ for $i'\le i$,
it is allowed to take $\mathbf g_\ell=p_\ell\mathbf u$ for $0\le\ell\le\delta_0-2$,
$\mathbf u_{i'}=p_{\delta_0-1+i'}\mathbf u$ for $i'\le i$,
and $\mathbf u_{i'}=\mathbf 0$ for $i'>i$ in \eqref{eq:linear-coefficient-evaluation}.
This then implies that $f_{\mathbf x}(X_{m+1})=u(\mathbf x)p(X_{m+1})$ for $\mathbf x\in U$
and $\mathbf c'=(u(\mathbf x)p(y))_{(\mathbf x,y)\in U\times\mathcal S_{m+1}}\in \mathscr C_{\mathcal S_{m+1}}(\mathbf E)$.
Note that the $i$th restricted coefficient of $\mathbf c'$ is $p_{\delta_0-1+i}\mathbf u=\mathbf u\notin F_i$.
By the uniqueness of the coefficients, $\mathbf c'\notin\mathscr C_{\mathcal S_{m+1}}(\mathbf F)$.
Since $u(\mathbf x)p(y)\ne0$ if and only if $u(\mathbf x)\ne0$ and $y\notin R$,
we have
$$\mathbf c'\in \mathscr C_{\mathcal S_{m+1}}(\mathbf E)\setminus\mathscr C_{\mathcal S_{m+1}}(\mathbf F)~~{\rm and}~~
d(\mathscr C_{\mathcal S_{m+1}}(\mathbf{E})\setminus\mathscr C_{\mathcal S_{m+1}}(\mathbf{F}))\leq \wt(\mathbf c')=(s+\delta_0-1-i)\eta_i.$$
This completes the proof of the bound \eqref{eq:coefficient-distance-bound}.

1) Fix $0\le j<s$. Since $j\le s-1$, we have $s+\delta_0-1-j\ge\delta_0$ and $\min\{\eta_k:\,k\le j\}\ge\min\{\eta_i:\,0\le i<s\}$.
It follows that
\[
 \begin{aligned}
 (s+\delta_0-1-j)\max\bigl\{d(E_j),~\min\{\eta_k:\,k\le j\}\bigr\}&\ge(s+\delta_0-1-j)d(E_j)\ge\min\{(s+\delta_0-1-i)d(E_i):\,0\le i<s\},~{\rm and}\\
 (s+\delta_0-1-j)\max\bigl\{d(E_j),~\min\{\eta_k:\,k\le j\}\bigr\}&\ge\delta_0\min\{\eta_k:\,k\le j\}\ge\delta_0\min\{\eta_i:\,0\le i<s\}.
 \end{aligned}
\]
Since the right-hand sides do not depend on $j$,
taking the minimum over $0\le j<s$ shows that
the lower bound in \eqref{eq:coefficient-distance-bound} is at least \eqref{eq.add.1}.

2) Since $E_j\setminus F_j\subseteq E_j\setminus\{\mathbf 0\}$,
we have $\eta_j\ge d(E_j)$ for $0\le j<s$. The described results follow from the two cases.
\begin{itemize}
  \item If $s=1$, then the lower and upper bounds in \eqref{eq:coefficient-distance-bound} are $\delta_0\max\{d(E_0),~\eta_0\}=\delta_0\eta_0$ and $\delta_0\eta_0$, respectively.

  \item If $E_i=E_0$ and $F_i=F_0$ for $0\le i<s$, then
$\max\bigl\{d(E_j),~\min\{\eta_k:\,k\le j\}\bigr\}=\eta_0$ for every $j$,
and both bounds in \eqref{eq:coefficient-distance-bound} equal $\min\{(s+\delta_0-1-j)\eta_0:\,0\le j<s\}=\delta_0\eta_0$.
\end{itemize}

3) If $\eta_i=d(E_i)$ for $0\le i<s$, we deduce from
$E_0\supseteq\cdots\supseteq E_{s-1}$ that $\eta_0\le\cdots\le\eta_{s-1}$.
This implies that $\max\bigl\{d(E_j),~\min\{\eta_k:\,k\le j\}\bigr\}=\eta_j$ for every $j$.
Therefore, both bounds in \eqref{eq:coefficient-distance-bound} equal $\min\{(s+\delta_0-1-j)\eta_j:\,0\le j<s\}$.
This completes the proof. 
\end{IEEEproof}
\medskip

We now take the nested codes $E_i$ and $F_i$ in Lemma~\ref{thm:coefficient-distance} to be specific GMCCs
and derive bounds on relative minimum distances via these nested pairs of GMCCs.
Theorem~\ref{thm:order-comparison} shows that these bounds can be exact and strictly exceed the known estimates in Lemma~\ref{lem:cartesian-relative-estimates}.
In $\Omega(\mathbf A)$, all monomials of $X_{m+1}$-degree at most $\delta_0-2$ are allowed, and the monomials of $X_{m+1}$-degree $\delta_0-1+i$ have their $\mathbf X$-exponents in $A_i$.
Hence, for $U=\mathcal X$, we have $\C(\mathbf A)=\mathscr C_{\mathcal S_{m+1}}\bigl((\C_{\mathcal X}(A_i))_{i=0}^{s-1}\bigr)$.

\begin{theorem}\label{cor:gmcc-distance-specialization}
Suppose that $\C_Z^\perp\subsetneq\C_X$, and let $\eta_i^X=d(\C_{\mathcal X}(A_i)\setminus\C_{\mathcal X}(D_i^B))$ and $\eta_i^Z=d(\C_{\mathcal X}(B_i)\setminus\C_{\mathcal X}(D_i^A))$ for $0\le i<s$. Then the
relative minimum distances of the CSS code $\mathcal Q$ in \eqref{eq:gmcc-css} satisfy
\begin{equation}\label{eq:relative-refined-lower}
 \begin{aligned}
 d_{XZ}&\ge\min\Biggl\{(s+\delta_0-1-i)\max\biggl\{\min\Bigl\{\prod_{j=1}^m(h_j-\alpha_j):\,\boldsymbol\alpha\in A_i\Bigr\},~\min\{\eta_k^X:\,k\le i\}\biggr\}:\,0\le i<s\Biggr\},\\
 d_{ZX}&\ge\min\Biggl\{(s+\delta_0-1-i)\max\biggl\{\min\Bigl\{\prod_{j=1}^m(h_j-\alpha_j):\,\boldsymbol\alpha\in B_i\Bigr\},~\min\{\eta_k^Z:\,k\le i\}\biggr\}:\,0\le i<s\Biggr\},
 \end{aligned}
\end{equation}
and, in particular,
\begin{equation}\label{eq:relative-lower-bounds}
 \begin{aligned}
 d_{XZ}&\ge\min\Bigl\{(s+\delta_0-1-i)\prod_{j=1}^m(h_j-\alpha_j):\,0\le i<s,~\boldsymbol\alpha\in A_i\Bigr\},\\
 d_{ZX}&\ge\min\Bigl\{(s+\delta_0-1-i)\prod_{j=1}^m(h_j-\alpha_j):\,0\le i<s,~\boldsymbol\alpha\in B_i\Bigr\}.
 \end{aligned}
\end{equation}
Moreover, for $0\le i<s$ and every $\boldsymbol\alpha\in T_i$,
\begin{equation}\label{eq:monomial-relative-upper}
 d_{XZ}\le(s+\delta_0-1-i)\prod_{j=1}^m(h_j-\alpha_j)~~{\rm and}~~d_{ZX}\le(\delta_0+i)\prod_{j=1}^m(\alpha_j+1).
\end{equation}
If $A_i=A_0$ and $D_i^B=D_0^B$ for $0\le i<s$, then $d_{XZ}=\delta_0\eta_0^X$. Similarly, if $B_i=B_0$ and $D_i^A=D_0^A$ for $0\le i<s$, then $d_{ZX}=\delta_0\eta_0^Z$.
\end{theorem}
\begin{IEEEproof}
Let $U=\mathcal X$. By Lemma~\ref{lem:gmcc-pair}.1), we have $\C_Z^\perp=\C(\mathbf D^B)\subsetneq\C_X=\C(\mathbf A)$.
Then $\C_X=\mathscr C_{\mathcal S_{m+1}}(\mathbf E)$ and $\C_Z^\perp=\mathscr C_{\mathcal S_{m+1}}(\mathbf F)$
for $E_i=\C_{\mathcal X}(A_i)$ and $F_i=\C_{\mathcal X}(D_i^B)$,
and hence $d_{XZ}=d(\mathscr C_{\mathcal S_{m+1}}(\mathbf E)\setminus\mathscr C_{\mathcal S_{m+1}}(\mathbf F))$.
Removing the common multiplier $\mathbf w$ from $\C_X^\perp\subsetneq\C_Z$ gives
$\C(\mathbf D^A)\subsetneq\C(\mathbf B)$ and
$d_{ZX}=d(\mathscr C_{\mathcal S_{m+1}}(\mathbf E)\setminus\mathscr C_{\mathcal S_{m+1}}(\mathbf F))$
for $E_i=\C_{\mathcal X}(B_i)$ and $F_i=\C_{\mathcal X}(D_i^A)$.
Since the chains $\mathbf A$ and $\mathbf B$ are nested, we have $E_0\supseteq\cdots\supseteq E_{s-1}$ in both cases.
By Lemma~\ref{lem:gmcc-pair}.2) and its proof,
we have $D_i^B\subseteq A_i$ and $D_i^A\subseteq B_i$ for $0\le i<s$, which gives $F_i\subseteq E_i$.
Moreover, the strict inclusions give $F_i\ne E_i$ for some $i$.
Hence, in both cases, the hypotheses of Lemma~\ref{thm:coefficient-distance} hold.
Since Lemma~\ref{lem:cartesian-duality-distance}.1) yields
$d(\C_{\mathcal X}(A_i))=\min\{\prod_{j=1}^m(h_j-\alpha_j):\,\boldsymbol\alpha\in A_i\}$ for nonempty $A_i$,
and similarly for $B_i$, the bounds \eqref{eq:relative-refined-lower} are
the lower bounds in \eqref{eq:coefficient-distance-bound} for these codes,
and \eqref{eq:relative-lower-bounds} follows from \eqref{eq.add.1}.

For a given $\boldsymbol\alpha\in T_i$, since $\alpha_j<h_j=|\mathcal S_j|$ for $1\le j\le m$
and $\delta_0-1+i<s+2\delta_0-2=|\mathcal S_{m+1}|$,
we can choose $R_j\subseteq\mathcal S_j$ with $|R_j|=\alpha_j$
and $R_{m+1}\subseteq\mathcal S_{m+1}$ with $|R_{m+1}|=\delta_0-1+i$.
Let
$$p(\mathbf{X},X_{m+1})=\prod_{j=1}^m\prod_{z\in R_j}(X_j-z)\prod_{y\in R_{m+1}}(X_{m+1}-y).$$
It is clear that every monomial $\mathbf X^{\boldsymbol\beta}X_{m+1}^{\ell}$ of $p$ satisfies $\boldsymbol\beta\le\boldsymbol\alpha$
coordinatewise and $\ell\le\delta_0-1+i$, and the coefficient of
$\mathbf{X}^{\boldsymbol\alpha} X_{m+1}^{\delta_0-1+i}$ in $p$ is one.
Then the monomials of $p$ with $\ell\le\delta_0-2$ lie in $\Omega(\mathbf A)$.
For a monomial $\mathbf X^{\boldsymbol\beta}X_{m+1}^{\ell}$ of $p$ with $\ell=\delta_0-1+i'$,
we have $i'\le i$ and $\boldsymbol\beta\le\boldsymbol\alpha\in A_i\subseteq A_{i'}$,
and since $A_{i'}$ is decreasing, it follows that $\boldsymbol\beta\in A_{i'}$.
Therefore, $\ev(p)\in\C(\mathbf A)$. Since the evaluations of the monomials
$\mathbf X^{\boldsymbol\beta}X_{m+1}^\ell$ with $(\boldsymbol\beta,\ell)\in\mathcal H$
form a basis of $\F_q^{n}$, the coefficient one of $\mathbf X^{\boldsymbol\alpha}X_{m+1}^{\delta_0-1+i}$
and $\boldsymbol\alpha\notin D_i^B$ give $\ev(p)\notin\C(\mathbf D^B)$.
We conclude that $\ev(p)\in \C_X\setminus\C_Z^\perp=\C(\mathbf A)\setminus \C(\mathbf{D}^B)$.
Since $p$ vanishes at $(\mathbf x,y)\in\mathcal P$ if and only if $x_j\in R_j$ for some $j$ or $y\in R_{m+1}$,
we have
$d_{XZ}\le\wt(\ev(p))=(s+\delta_0-1-i)\prod_{j=1}^m(h_j-\alpha_j).$

For the second bound on $d_{ZX}$, since
$\boldsymbol\alpha\in A_i\cap\rho(B_{s-1-i})$, we have $\rho(\boldsymbol\alpha)\in B_{s-1-i}\setminus\rho(\Lambda\setminus A_i)
=B_{s-1-i}\setminus D_{s-1-i}^A$.
Note that $\delta_0-1+(s-1-i)=s+\delta_0-2-i$.
Let $p'$ be the polynomial defined by using $\mathbf X^{\rho(\boldsymbol\alpha)}X_{m+1}^{s+\delta_0-2-i}$ in place of
$\mathbf X^{\boldsymbol\alpha}X_{m+1}^{\delta_0-1+i}$ in $p$.
A similar argument as above shows that $\ev(p')\diag(\mathbf{w}) \in \C_Z\setminus\C_X^\perp$.
Since $(s+2\delta_0-2)-(s+\delta_0-2-i)=\delta_0+i$ and
$h_j-(h_j-1-\alpha_j)=\alpha_j+1$,
we deduce that
$d_{ZX}\leq \wt(\ev(p')\diag(\mathbf w))=(\delta_0+i)\prod_{j=1}^m(\alpha_j+1).$
This proves the upper bounds \eqref{eq:monomial-relative-upper}.

Finally, the two equalities for constant nested pairs directly follow from Lemma~\ref{thm:coefficient-distance}.2).
\end{IEEEproof}
\medskip

Next, we compare the bounds in Theorem~\ref{cor:gmcc-distance-specialization} with the two known estimates in Lemma~\ref{lem:cartesian-relative-estimates}.
The following family shows that our lower bounds \eqref{eq:relative-refined-lower} can be exact and exceed both of them. Let $m=2$, $M\ge3$, $q\ge\max\{M,~4\}$, $h_1=M$, $h_2=3$, and $s=\delta_0=2$. We take
\begin{align}\label{eq.add2}
   A_0=A_1=B_0=B_1=\{0,\ldots,M-1\}\times\{0\}\cup\{0,\ldots,M-2\}\times\{1\}.
\end{align}

\begin{theorem}\label{thm:order-comparison}
The CSS code $\mathcal Q$ in \eqref{eq:gmcc-css} with the chains \eqref{eq.add2}
is impure and has parameters $\dsb{12M,2M-4,8}_q$.
Moreover, for the nested pair $\C_Z^\perp\subsetneq\C_X$, both estimates in Lemma~\ref{lem:cartesian-relative-estimates} are at most six for every monomial order.
\end{theorem}
\begin{IEEEproof}
By \eqref{eq:dual-chains} and $M\geq 3$, we have
$$D_0^A=D_1^A=D_0^B=D_1^B=\{0,\ldots,M-1\}\times\{0\}\cup\{(0,1)\}
~~{\rm and}~~T_0=T_1=\{1,\ldots,M-2\}\times\{1\}\neq \varnothing.$$
Since $D_i^B\subseteq A_i$ for $i=0,~1$, it follows from
Lemmas~\ref{lem:gmcc-pair}.2) and \ref{lem:gmcc-pair}.3) that $\C_Z^\perp\subsetneq\C_X$
and $\mathcal Q$ has parameters $\dsb{12M,2M-4,d}_q$ with minimum distance
$d=\min\{d_{XZ},~d_{ZX}\}$ and minimum stabilizer weight $\sigma=\min\{4,~3\cdot3,~2\cdot3\}=4$,
where the latter value follows from \eqref{eq:multivariate-stabilizer} and the fact that
$\prod_{j=1}^2(h_j-\alpha_j)\ge3$ for $\boldsymbol\alpha\in D_i^A=D_i^B$ and $i=0,~1$, with equality at $\boldsymbol\alpha=(M-1,0)$.
According to Theorem~\ref{cor:gmcc-distance-specialization},
it suffices to compute $\eta_0^X$ and $\eta_0^Z$ for determining $d_{XZ}$ and $d_{ZX}$, respectively.

Note that $\C_{\mathcal X}(A_0)$ consists of the evaluations of $\psi_0(X_1)+X_2\psi_1(X_1)$
for $\psi_0,~\psi_1\in\F_q[X_1]$ with $\deg(\psi_0)\le M-1$ and $\deg(\psi_1)\le M-2$,
and $\C_{\mathcal X}(D_0^B)$ consists of those with $\deg(\psi_0)\le M-1$ and $\psi_1\in\F_q$.
Every vector in $\C_{\mathcal X}(A_0)\setminus\C_{\mathcal X}(D_0^B)$ is the evaluation of
$\psi_0(X_1)+X_2\psi_1(X_1)$ with a nonconstant polynomial $\psi_1$ of degree at most $M-2$.
The polynomial $\psi_1$ is nonzero at two or more points of $\mathcal S_1$,
and at each such point $x_1$, the polynomial $\psi_0(x_1)+X_2\psi_1(x_1)$ in $X_2$
has degree one and hence is nonzero at two or more points of $\mathcal S_2$.
Since the evaluation of $(X_2-\xi)\prod_{z\in R}(X_1-z)$ with $\xi\in\mathcal S_2$ and
$R\subseteq\mathcal S_1$ of size $M-2$ lies in $\C_{\mathcal X}(A_0)\setminus\C_{\mathcal X}(D_0^B)$
and has weight four, we deduce that $\eta_0^X=4$.
Since $B_0=A_0$ and $D_0^A=D_0^B$,
the nested pair $\C_{\mathcal X}(D_0^A)\subsetneq\C_{\mathcal X}(B_0)$ defining $\eta_0^Z$
is the nested pair $\C_{\mathcal X}(D_0^B)\subsetneq\C_{\mathcal X}(A_0)$ defining $\eta_0^X$,
and hence $\eta_0^Z=\eta_0^X=4$.
With Theorem~\ref{cor:gmcc-distance-specialization}, we conclude that
$d=\delta_0\eta_0^X=8>4=\sigma$.
Therefore, $\mathcal Q$ is an impure $\dsb{12M,2M-4,8}_q$ CSS code.

To obtain the estimates in Lemma~\ref{lem:cartesian-relative-estimates},
we identify the defining sets $\Omega(\mathbf A)$ of $\C_X$ and $\Omega(\mathbf D^B)$ of $\C_Z^\perp$
with sets of monomials in $X_1,~X_2,~X_3$.
By Lemma~\ref{lem:gmcc-pair}.1), we have $\C_X=\C_{\mathcal P}(\Omega(\mathbf A))$ and
$\C_Z^\perp=\C_{\mathcal P}(\Omega(\mathbf D^B))$, and hence $\C_Z=\C_{\mathcal P}(\Omega(\mathbf D^B))^\perp$ and $\C_X^\perp=\C_{\mathcal P}(\Omega(\mathbf A))^\perp$.
For a monomial order $\prec$, Lemma~\ref{lem:cartesian-relative-estimates} with $\Omega_1=\Omega(\mathbf A)$ and $\Omega_2=\Omega(\mathbf D^B)$ gives
\[
 \begin{aligned}
 d_{XZ}=d(\C_{\mathcal P}(\Omega(\mathbf A))\setminus\C_{\mathcal P}(\Omega(\mathbf D^B)))&\ge\min\Bigl\{(M-i_1)(3-i_2)(4-i_3):\,\mathbf i\in\Omega(\mathbf A),~\mathbf i\succeq\min\nolimits_\prec(\Omega(\mathbf A)\setminus\Omega(\mathbf D^B))\Bigr\},~{\rm and}\\
 d_{ZX}=d(\C_{\mathcal P}(\Omega(\mathbf D^B))^\perp\setminus\C_{\mathcal P}(\Omega(\mathbf A))^\perp)&\ge\min\Bigl\{(i_1+1)(i_2+1)(i_3+1):\,\mathbf i\in\mathcal H\setminus\Omega(\mathbf D^B),~\mathbf i\preceq\max\nolimits_\prec(\Omega(\mathbf A))\Bigr\}.
 \end{aligned}
\]

Note that $\Omega(\mathbf A)\setminus\Omega(\mathbf D^B)=
\{X_1^{\nu}X_2X_3^k:\,1\le\nu\le M-2,~k=1,~2\}$ and every element of this set is divisible by $X_1X_2X_3$.
Hence, $\min_\prec(\Omega(\mathbf A)\setminus\Omega(\mathbf D^B))=X_1X_2X_3$ for every monomial order.
Since $X_1^{M-1}X_2^2\cdot X_1^{M-1}X_3^2=X_1^{2M-4}(X_1X_2X_3)^2\succeq(X_1X_2X_3)^2$,
one of the monomials $X_1^{M-1}X_2^2,~X_1^{M-1}X_3^2\in\Omega(\mathbf A)$ does not precede $X_1X_2X_3$.
Hence, for every monomial order, at least one of the exponents $(M-1,2,0)$ and $(M-1,0,2)$ lies in the set over which the first minimum is taken. The corresponding values of $(M-i_1)(3-i_2)(4-i_3)$ are $1\cdot1\cdot4=4$ and $1\cdot3\cdot2=6$, respectively, and it follows that the first minimum is at most six.
Similarly, $X_1^{M-2}X_2X_3^2\in\Omega(\mathbf A)$ gives $X_1^{M-2}X_2X_3^2\preceq\max_\prec(\Omega(\mathbf A))$, and $(X_1^{M-2}X_2X_3^2)^2=X_1^{2M-4}\cdot X_2^2X_3\cdot X_3^3$. Hence, one of the monomials $X_2^2X_3,~X_3^3\notin\Omega(\mathbf D^B)$ does not succeed $\max_\prec(\Omega(\mathbf A))$.
It follows that at least one of the exponents $(0,2,1)$ and $(0,0,3)$ lies in the set over which
the second minimum is taken. The corresponding values of $(i_1+1)(i_2+1)(i_3+1)$
are $1\cdot3\cdot2=6$ and $1\cdot1\cdot4=4$, respectively, and hence the second minimum is also at most six.
As a result, both estimates in Lemma~\ref{lem:cartesian-relative-estimates} are at most six for every monomial order,
whereas our bounds give the exact value $d_{XZ}=d_{ZX}=8$.
\end{IEEEproof}

\subsection{Recovering sets and minimum localities of $\mathcal{Q}$}\label{subsec:gmcc-locality}

We now study recovering sets of the CSS code $\mathcal Q$ in \eqref{eq:gmcc-css}.
We call $J_{\mathbf{x}}=\{\mathbf{x}\}\times\mathcal S_{m+1}$ a
\emph{block}.  The blocks satisfy
\begin{equation}\label{eq:disjoint-recovering-partition}
 \mathcal P=\bigsqcup_{\mathbf{x}\in\mathcal X}J_{\mathbf{x}}~~{\rm and}~~
 |J_{\mathbf{x}}|=s+2\delta_0-2,
\end{equation}
and the following lemma shows that they form disjoint $(s+\delta_0-1,\delta_0)$-recovering sets of
$\mathcal{Q}$, and gives conditions for $\mathcal{Q}$ to be GKT-optimal.
\begin{lemma}\label{lem:blocks-recovering}
Suppose that $\C_Z^\perp\subsetneq\C_X$, and let $\mathcal Q$ be the CSS code in \eqref{eq:gmcc-css} with parameters $\dsb{n,\kappa,d}_q$. Then the following results hold.
\begin{enumerate}[label=\arabic*),leftmargin=*,itemsep=0pt,topsep=1pt,beginpenalty=10000]
\item Every block $J_{\mathbf{x}}$ in \eqref{eq:disjoint-recovering-partition} is an $(s+\delta_0-1,\delta_0)$-recovering set of $\mathcal Q$.
Moreover, $\mathcal Q$ has disjoint $(s+\delta_0-1,\delta_0)$-recovering sets with $J(\mathbf{x},y)=J_{\mathbf{x}}$ for every coordinate $(\mathbf{x},y)$.

\item If every $A_i$ and $B_i$ is nonempty, then $P_{J_{\mathbf{x}}}(\C_X)$ and $P_{J_{\mathbf{x}}}(\C_Z)$ are
$[s+2\delta_0-2,s+\delta_0-1,\delta_0]_q$ MDS codes.
\item With the disjoint $(s+\delta_0-1,\delta_0)$-recovering sets in 1),
$\mathcal Q$ is GKT-optimal if and only if
\begin{equation}\label{eq:gkt-blocks-optimal}
 \kappa>s\prod_{j=1}^mh_j-2\left(d-(\delta_0-1)\left\lceil\frac{d}{s+\delta_0-1}\right\rceil\right).
\end{equation}
In particular, if $d=c(s+\delta_0-1)$ for an integer $c\ge1$,
then $\mathcal Q$ is GKT-optimal if and only if
$\kappa>s\prod_{j=1}^mh_j-2cs$.
\end{enumerate}
\end{lemma}
\begin{IEEEproof}
1) Fix $\mathbf x\in\mathcal X$, and let $(r,\delta)=(s+\delta_0-1,\delta_0)$. Since $|J_{\mathbf x}|=s+2\delta_0-2=r+\delta-1\ge\delta$, it suffices by Lemma~\ref{lem:css-recovery-criterion} to show that \eqref{eq:css-recovery-criterion} holds for $J=J_{\mathbf x}$ and every $I\subseteq J_{\mathbf{x}}$ with $|I|=\delta_0-1$. We show that all four codes in \eqref{eq:css-recovery-criterion} are zero.
After removing the nonzero multipliers, every codeword of $P_{J_{\mathbf{x}}}(\C_X)$ or of $P_{J_{\mathbf{x}}}(\C_Z)$ is the evaluation on $\mathcal S_{m+1}$ of a polynomial in $X_{m+1}$ of degree at most $s+\delta_0-2$.
If such a codeword is supported in $I$, then this polynomial
vanishes at the $s+2\delta_0-2-|I|=s+\delta_0-1>s+\delta_0-2$ coordinates of $J_{\mathbf x}$ outside $I$,
implying that the polynomial is zero. Hence,
$S_I(P_{J_{\mathbf{x}}}(\C_X))=S_I(P_{J_{\mathbf{x}}}(\C_Z))=\{\mathbf{0}\}$.
Since $\C_Z^\perp\subseteq\C_X$, every codeword of $\C_Z^\perp$ supported in $I$ is
a codeword of $\C_X$ supported in $I\subseteq J_{\mathbf x}$,
which gives $S_I(\C_Z^\perp)\subseteq S_I(P_{J_{\mathbf{x}}}(\C_X))=\{\mathbf{0}\}$.
Similarly, $\C_X^\perp\subseteq\C_Z$ gives $S_I(\C_X^\perp)=\{\mathbf 0\}$.
Hence, both equalities in \eqref{eq:css-recovery-criterion} hold.
With Lemma~\ref{lem:css-recovery-criterion}, $J_{\mathbf x}$ is an $(s+\delta_0-1,\delta_0)$-recovering set of $\mathcal Q$.

Moreover, it is clear from \eqref{eq:disjoint-recovering-partition} that these blocks are pairwise disjoint and
every coordinate $(\mathbf x,y)\in\mathcal P$ lies in the block $J_{\mathbf x}$ with $|J_{\mathbf x}|=r+\delta-1$.
Taking $J(\mathbf x,y)=J_{\mathbf x}$ in Definition~\ref{def:recovering-sets}, we conclude that $\mathcal Q$ has disjoint $(s+\delta_0-1,\delta_0)$-recovering sets.

2) Since every $A_i$ and $B_i$ is nonempty, all of them contain $\mathbf{0}$. In this case, $P_{J_{\mathbf{x}}}(\C_X)$ and $P_{J_{\mathbf{x}}}(\C_Z)$ consist of the evaluations of all polynomials in $X_{m+1}$ of degree at most $s+\delta_0-2$,
with nonzero column multipliers for $\C_Z$.
A nonzero such polynomial has at most $s+\delta_0-2$ roots, and hence its evaluation has weight
at least $(s+2\delta_0-2)-(s+\delta_0-2)=\delta_0$.
It follows that both codes have dimension $s+\delta_0-1$ and minimum distance at least $\delta_0$.
By the well-known Singleton bound, they are $[s+2\delta_0-2,s+\delta_0-1,\delta_0]_q$ MDS codes.

3) By 1), the GKT bound \eqref{eq:gkt-endpoint} can be applied to
$\mathcal Q$ with $n=(s+2\delta_0-2)\prod_{j=1}^mh_j$
and $(r,\delta)=(s+\delta_0-1,\delta_0)$.
Since $r+\delta-1=s+2\delta_0-2$, we have $\bigl(1-2(\delta-1)/(r+\delta-1)\bigr)n=s\prod_{j=1}^mh_j$,
and \eqref{eq:gkt-endpoint} reads
\begin{equation}\label{eq:gkt-blocks}
 \kappa\le s\prod_{j=1}^mh_j-2\left(d-1-(\delta_0-1)\left\lceil\frac{d-1}{s+\delta_0-1}\right\rceil\right).
\end{equation}
Since $\mathcal Q$ satisfies \eqref{eq:gkt-blocks}, and \eqref{eq:gkt-blocks}
with $d$ replaced by $d+1$ is exactly the negation of \eqref{eq:gkt-blocks-optimal},
we immediately conclude that $\mathcal Q$ is GKT-optimal
if and only if \eqref{eq:gkt-blocks-optimal} holds by definition.
The last assertion follows from this by a straightforward calculation.
\end{IEEEproof}
\medskip

Based on Theorem~\ref{lem:css-support-restrictions},
the $(r,\delta)$-recovering sets of $\mathcal Q$ are determined by the local relative distances
in \eqref{eq:general-local-relative-distance}. For a union of blocks $U\times\mathcal S_{m+1}$
with $\varnothing\ne U\subseteq\mathcal X$, these involve punctured and shortened codes of
length $(s+2\delta_0-2)|U|$ in the $m+1$ variables.
The following lemma expresses them through the nested pairs
$S_U(\C_{\mathcal X}(D_i^B))\subseteq P_U(\C_{\mathcal X}(A_i))$ of length $|U|$ in
only the $m$ variables $\mathbf X$, which makes the bounds and the equality cases of
Lemma~\ref{thm:coefficient-distance} applicable to $\lambda(U\times\mathcal S_{m+1})$.
\begin{lemma}\label{lem:block-relative-distance}
Suppose that $\C_Z^\perp\subsetneq\C_X$, and let $\varnothing\ne U\subseteq\mathcal X$. For $0\le i<s$, let either $E_i=P_U(\C_{\mathcal X}(A_i))$ and $F_i=S_U(\C_{\mathcal X}(D_i^B))$, or $E_i=P_U(\C_{\mathcal X}(B_i))$ and $F_i=S_U(\C_{\mathcal X}(D_i^A))$. Then the following results hold.
\begin{enumerate}[label=\arabic*),leftmargin=*,itemsep=0pt,topsep=1pt,beginpenalty=10000]
\item $E_0\supseteq\cdots\supseteq E_{s-1}$ and $F_i\subseteq E_i$ for $0\le i<s$.
\item In the first case, $P_{U\times\mathcal S_{m+1}}(\C_X)=\mathscr C_{\mathcal S_{m+1}}(\mathbf E)$ and $S_{U\times\mathcal S_{m+1}}(\C_Z^\perp)=\mathscr C_{\mathcal S_{m+1}}(\mathbf F)$. In the second case, $P_{U\times\mathcal S_{m+1}}(\C(\mathbf B))=\mathscr C_{\mathcal S_{m+1}}(\mathbf E)$ and $S_{U\times\mathcal S_{m+1}}(\C(\mathbf D^A))=\mathscr C_{\mathcal S_{m+1}}(\mathbf F)$.
\end{enumerate}
\end{lemma}
\begin{IEEEproof}
1) By Lemmas~\ref{lem:gmcc-pair}.1) and \ref{lem:gmcc-pair}.2), we have $\C_Z^\perp=\C(\mathbf D^B)$ and $D_i^B\subseteq A_i$ for $0\le i<s$.
The chain $A_0\supseteq\cdots\supseteq A_{s-1}$ gives $E_0\supseteq\cdots\supseteq E_{s-1}$, and $D_i^B\subseteq A_i$ gives $F_i\subseteq P_U(\C_{\mathcal X}(D_i^B))\subseteq E_i$.
Similarly, removing the common multiplier $\mathbf w$ from $\C_X^\perp\subseteq\C_Z$ gives $\C(\mathbf D^A)\subseteq\C(\mathbf B)$, that is, $D_i^A\subseteq B_i$ for $0\le i<s$, and the chain $B_0\supseteq\cdots\supseteq B_{s-1}$ then gives the second case.

2) Restricting a codeword $(f_{\mathbf x}(y))_{(\mathbf x,y)\in\mathcal X\times\mathcal S_{m+1}}$ of
$\C_X=\mathscr C_{\mathcal S_{m+1}}((\C_{\mathcal X}(A_i))_{i=0}^{s-1})$ to $U\times\mathcal S_{m+1}$
also restricts its coefficient vectors to $U$.
As the codeword ranges over $\C_X$, the restrictions of its lower coefficients to $U$ range over $\F_q^{|U|}$
and the restriction of its $i$th restricted coefficient to $U$ ranges over $P_U(\C_{\mathcal X}(A_i))=E_i$.
This yields $P_{U\times\mathcal S_{m+1}}(\C_X)=\mathscr C_{\mathcal S_{m+1}}(\mathbf E)$.
Since $\deg_{X_{m+1}}(f_{\mathbf x})<s+2\delta_0-2$, a codeword of
$\C(\mathbf D^B)=\mathscr C_{\mathcal S_{m+1}}((\C_{\mathcal X}(D_i^B))_{i=0}^{s-1})$
is supported in $U\times\mathcal S_{m+1}$ if and only if its lower coefficients and
its restricted coefficients are supported in $U$.
After restriction to $U\times\mathcal S_{m+1}$, the lower coefficients of
these codewords then range over $\F_q^{|U|}$ and their $i$th restricted coefficients
range over $S_U(\C_{\mathcal X}(D_i^B))=F_i$.
As a result, $S_{U\times\mathcal S_{m+1}}(\C_Z^\perp)=\mathscr C_{\mathcal S_{m+1}}(\mathbf F)$.

For the second case, it follows from Lemma~\ref{lem:gmcc-pair}.1) that
$\C_Z=\C(\mathbf B)\diag(\mathbf w)$ and $\C_X^\perp=\C(\mathbf D^A)\diag(\mathbf w)$.
Since multiplication by $\diag(\mathbf w)$ preserves the supports and weights of any codeword,
the desired identities in this case hold by using an argument similar to the first case.
\end{IEEEproof}
\medskip

Note that, when $\sigma=s+\delta_0$, every subset of a block of size $s+\delta_0$ is a minimal support of $\C_X^\perp$ and of $\C_Z^\perp$.
From Theorem~\ref{lem:css-support-restrictions}.2), these minimal supports force every recovering set to be of
the form $U\times\mathcal S_{m+1}$ with $U\subseteq\mathcal X$ when $\delta>s+\delta_0-1$.
The following theorem is based on this observation.
It determines $r_{\mathcal Q}^{\min}(\delta)$ for any $2\le\delta\le\delta_0$ via Lemma~\ref{lem:blocks-recovering},
and reduces the case $\delta>s+\delta_0-1$ to the local relative distances $\lambda(U\times\mathcal S_{m+1})$ in Lemma~\ref{lem:block-relative-distance}.
Here and below, we say that $r_{\mathcal Q}^{\min}(\delta)$ is {\em attained by} a family of sets if these sets are $(r_{\mathcal Q}^{\min}(\delta),\delta)$-recovering sets of $\mathcal Q$ that cover all coordinates, and that it is {\em attained exactly by} this family if, in addition, these sets are the only $(r_{\mathcal Q}^{\min}(\delta),\delta)$-recovering sets of $\mathcal Q$.

\begin{theorem}\label{thm:all-s1}
Suppose that $\C_Z^\perp\subsetneq\C_X$ and that the CSS code $\mathcal Q$ in \eqref{eq:gmcc-css} has
minimum stabilizer weight $\sigma=s+\delta_0$, and let $r\ge1$ and $\delta\ge2$ be integers.
Then the following results hold.
\begin{enumerate}[label=\arabic*),leftmargin=*,itemsep=0pt,topsep=1pt,beginpenalty=10000]
\item If $2\le\delta\le\delta_0$, then
$$r_{\mathcal Q}^{\min}(\delta)=s+\delta_0-1,$$
attained by the sets $J\subseteq J_{\mathbf{x}}$ with $\mathbf x\in\mathcal X$ and $|J|=s+\delta_0+\delta-2$,
where $J_{\mathbf x}$ is the block in \eqref{eq:disjoint-recovering-partition}.
Moreover, if $s=1$, then $r_{\mathcal Q}^{\min}(\delta)$ is attained exactly by these sets.

\item Let $\delta>s+\delta_0-1$, and suppose that there is an integer $\gamma_0$ with
$1\le\gamma_0\le |\mathcal X|$ and $(s+2\delta_0-2)\gamma_0\ge\delta$ such that,
for every $\varnothing\ne U\subseteq\mathcal X$, $\lambda(U\times\mathcal S_{m+1})\ge\delta$
if and only if $|U|\ge\gamma_0$.
Then
$$r_{\mathcal Q}^{\min}(\delta)=(s+2\delta_0-2)\gamma_0-\delta+1,$$
attained exactly by the sets $U\times\mathcal S_{m+1}$ with $U\subseteq\mathcal X$ and $|U|=\gamma_0$. Moreover, every $(r,\delta)$-recovering set of $\mathcal Q$ is
$U\times\mathcal S_{m+1}$ for some $U\subseteq\mathcal X$ with $|U|\ge\gamma_0$.
\end{enumerate}
\end{theorem}
\begin{IEEEproof}
Let $\mathbf x\in\mathcal X$ and $\mathcal{K}\subseteq J_{\mathbf x}$ with
$|\mathcal{K}|=s+\delta_0$, and let
$p(X_{m+1})=\prod_{(\mathbf x,y)\in J_{\mathbf{x}}\setminus \mathcal{K}}(X_{m+1}-y)$
be a polynomial of degree $(s+2\delta_0-2)-(s+\delta_0)=\delta_0-2$.
Taking $f_{\mathbf x}=p$ and $f_{\mathbf x'}=0$ for $\mathbf x'\ne\mathbf x$ in
\eqref{eq:linear-coefficient-evaluation} then gives codewords of
$\C(\mathbf D^A)$ and $\C(\mathbf D^B)$ with support $\mathcal{K}$.
By Lemma~\ref{lem:gmcc-pair}.1), we have $\C_X^\perp=\C(\mathbf D^A)\diag(\mathbf w)$ and $\C_Z^\perp=\C(\mathbf D^B)$, and hence both dual codes contain a codeword of weight $s+\delta_0$ with support $\mathcal{K}$. Since $\sigma=s+\delta_0$, it follows that $d(\C_X^\perp)=d(\C_Z^\perp)=s+\delta_0$, and hence $\mathcal{K}$ is a minimal support of $\C_X^\perp$ and of $\C_Z^\perp$.

1) Assume that $A_{s-1}=\varnothing$.
Then $D_0^A=\Lambda$ and $(\boldsymbol\tau,\delta_0-1)\in\Omega(\mathbf D^A)$.
By Lemma~\ref{lem:cartesian-duality-distance}.1), we obtain
$d(\C_X^\perp)\le(s+2\delta_0-2)-(\delta_0-1)=s+\delta_0-1$, a contradiction.
Similarly, $B_{s-1}\ne\varnothing$.
This implies that all sets $A_i,~B_i$ are nonempty and
$P_{J_{\mathbf{x}}}(\C_X)$ and $P_{J_{\mathbf{x}}}(\C_Z)$ are $[s+2\delta_0-2,s+\delta_0-1,\delta_0]_q$ MDS codes
by Lemma~\ref{lem:blocks-recovering}.2).
Then, for every set $J\subseteq J_{\mathbf x}$ with $|J|=s+\delta_0+\delta-2$,
it is easy to check that the punctured codes $P_J(\C_X)$ and $P_J(\C_Z)$ are MDS codes of minimum distance $\delta$.
Every nonzero codeword of $\C_X$ or of $\C_Z$ has a nonzero restriction to some $J_{\mathbf x}$,
which is a codeword of one of these MDS codes.
By Lemma~\ref{lem:css-construction}, we deduce that
$$d\ge\min\{d(\C_X),~d(\C_Z)\}\ge\delta_0=\sigma-s\ge\delta\geq 2.$$
Since $\delta\le\delta_0<s+\delta_0=\sigma$, Corollary~\ref{cor:locality-local-distance}.2) shows that these sets are quantum $(s+\delta_0-1,\delta)$-recovering.
Since these sets cover all coordinates, $r_{\mathcal Q}^{\min}(\delta)\le s+\delta_0-1$.

To prove $r_{\mathcal Q}^{\min}(\delta)\geq s+\delta_0-1$, we let
$J$ be a quantum $(r,\delta)$-recovering set.
Since $\delta\le\delta_0<\sigma$, Corollary~\ref{cor:locality-local-distance}.2) shows that $J$ is a classical $(r,\delta)$-recovering set of $\C_X$, that is, $d(P_J(\C_X))\ge\delta$.
Since $\C_X$ contains the evaluation of the constant polynomial $1$, which has full support,
this code is nonzero.
By the well-known Singleton bound, $\dim(P_J(\C_X)^\perp)=|J|-\dim(P_J(\C_X))\ge\delta-1$.
With \cite[Theorem~1.5.7(i)]{HP2003}, we obtain $P_J(\C_X)^\perp=S_J(\C_X^\perp)$.
Hence, $d(P_J(\C_X)^\perp)\ge d(\C_X^\perp)=s+\delta_0$ and
$P_J(\C_X)^\perp$ is nonzero since its dimension is at least $\delta-1\ge1$.
Applying the Singleton bound to $P_J(\C_X)^\perp$ then gives
$$|J|\ge d(P_J(\C_X)^\perp)+\dim(P_J(\C_X)^\perp)-1\ge (s+\delta_0)+(\delta-1)-1=s+\delta_0+\delta-2.$$
Since $|J|\le r+\delta-1$ by Definition~\ref{def:quantum-locality},
we have $r+\delta-1\ge s+\delta_0+\delta-2$, that is, $r\ge s+\delta_0-1$.
Hence, $\mathcal Q$ has no $(r,\delta)$-recovering set for any $r<s+\delta_0-1$,
and in particular no $(r,\delta)$-locality. Since $r_{\mathcal Q}^{\min}(\delta)$ is the smallest $r$
for which $\mathcal Q$ has $(r,\delta)$-locality,
$r_{\mathcal Q}^{\min}(\delta)\ge s+\delta_0-1$ holds immediately.
In summary, $r_{\mathcal Q}^{\min}(\delta)=s+\delta_0-1$.

We now fix $s=1$ and let $J$ be a $(\delta_0,\delta)$-recovering set.
By Definition~\ref{def:quantum-locality} and the above deductions, we have
$|J|=\delta_0+\delta-1\le2\delta_0-1$. Suppose that $J$ meets at least two blocks.
Since $|J|\le2\delta_0-1$, there is $\mathbf x\in\mathcal X$ with $1\le|J\cap J_{\mathbf x}|\le\delta_0-1$.
Fix $(\mathbf x,z)\in J\cap J_{\mathbf x}$. By Lagrange interpolation,
we can find a polynomial $p\in\F_q[X_{m+1}]$ of degree at most
$|J\cap J_{\mathbf x}|-1\le\delta_0-2$ with $p(z)=1$ and $p(y)=0$
for every $(\mathbf x,y)\in(J\cap J_{\mathbf x})\setminus\{(\mathbf x,z)\}$.
Since the lower coefficients are unrestricted, taking $f_{\mathbf x}=p$ and $f_{\mathbf x'}=0$
for $\mathbf x'\ne\mathbf x$ in \eqref{eq:linear-coefficient-evaluation} provides
a codeword of $P_J(\C_X)$ of weight one, which contradicts $d(P_J(\C_X))\ge\delta\ge2$.
Hence, $J\subseteq J_{\mathbf x}$ for some $\mathbf x\in\mathcal X$
and $r_{\mathcal Q}^{\min}(\delta)$ is attained exactly
by the sets $J\subseteq J_{\mathbf x}$ with $\mathbf x\in\mathcal X$ and $|J|=\delta_0+\delta-1$.

2)
We first show that every $(r,\delta)$-recovering set $J$ has the form $U\times\mathcal S_{m+1}$ with
$\varnothing\ne U\subseteq\mathcal X$. Suppose that $J\cap J_{\mathbf{x}}\ne\varnothing$ and
$J_{\mathbf x}\not\subseteq J$ for some $\mathbf x\in\mathcal X$.
Then there exist $(\mathbf x,y)\in J\cap J_{\mathbf x}$ and $(\mathbf x,y')\in J_{\mathbf x}\setminus J$. Since $2\le s+\delta_0\le s+2\delta_0-2$, we can then choose
$\mathcal{K}\subseteq J_{\mathbf{x}}$ with $|\mathcal{K}|=s+\delta_0$
and $(\mathbf x,y),~(\mathbf x,y')\in\mathcal{K}$.
Hence, $\mathcal{K}\cap J\ne\varnothing$
and $\mathcal{K}\not\subseteq J$.
Since $\mathcal{K}$ is a minimal support of $\C_Z^\perp$ with $|\mathcal{K}|=s+\delta_0\le\delta$,
this contradicts Theorem~\ref{lem:css-support-restrictions}.2).
Hence, for every $\mathbf x\in\mathcal X$, either $J\cap J_{\mathbf x}=\varnothing$ or $J_{\mathbf x}\subseteq J$, which gives $J=U\times\mathcal S_{m+1}$ with $U=\{\mathbf x\in\mathcal X:\,J_{\mathbf x}\subseteq J\}\ne\varnothing$. By Theorem~\ref{lem:css-support-restrictions}, a set $U\times\mathcal S_{m+1}$ with $\varnothing\ne U\subseteq\mathcal X$ is an $(r,\delta)$-recovering set if and only if $\delta\le(s+2\delta_0-2)|U|\le r+\delta-1$ and $\lambda(U\times\mathcal S_{m+1})\ge\delta$.

Combining with the hypothesis on $\gamma_0$, it follows that every $(r,\delta)$-recovering set is $U\times\mathcal S_{m+1}$ with $|U|\ge\gamma_0$ and $(s+2\delta_0-2)|U|\le r+\delta-1$. Hence, $r\ge (s+2\delta_0-2)\gamma_0-\delta+1$. Conversely, let $r\ge (s+2\delta_0-2)\gamma_0-\delta+1$. For every $U$ with $|U|=\gamma_0$, we have $\delta\le (s+2\delta_0-2)\gamma_0\le r+\delta-1$ and $\lambda(U\times\mathcal S_{m+1})\ge\delta$, and hence $U\times\mathcal S_{m+1}$ is $(r,\delta)$-recovering. Since $1\le\gamma_0\le |\mathcal X|$, these sets cover all coordinates. It follows that $r_{\mathcal Q}^{\min}(\delta)=(s+2\delta_0-2)\gamma_0-\delta+1$.

For $r=r_{\mathcal Q}^{\min}(\delta)$, the inequality $(s+2\delta_0-2)|U|\le r+\delta-1$ forces $|U|=\gamma_0$, and hence the $(r_{\mathcal Q}^{\min}(\delta),\delta)$-recovering sets are exactly the sets $U\times\mathcal S_{m+1}$ with $|U|=\gamma_0$.
\end{IEEEproof}

The following corollary shows that $\sigma=\delta_0+1$ when $s=1$,
and hence the two ranges of $\delta$ in Theorem~\ref{thm:all-s1} cover every $2\le\delta\le d$.
It also determines $\lambda(U\times\mathcal S_{m+1})$.
\begin{corollary}\label{cor:s1-specialization}
Let $s=1$, and suppose that $\C_Z^\perp\subsetneq\C_X$. Then the CSS code $\mathcal Q$ in \eqref{eq:gmcc-css}
has minimum stabilizer weight $\sigma=\delta_0+1$ and
\begin{equation}\label{eq:s1-block-distance}
 \lambda(U\times\mathcal S_{m+1})=\delta_0\min\bigl\{d(P_U(\C_{\mathcal X}(A_0))\setminus S_U(\C_{\mathcal X}(D_0^B))),~d(P_U(\C_{\mathcal X}(B_0))\setminus S_U(\C_{\mathcal X}(D_0^A)))\bigr\}
\end{equation}
for every nonempty $U\subseteq\mathcal X$.
\end{corollary}
\begin{IEEEproof}
By Lemma~\ref{lem:gmcc-pair}.2), we have $T_0=A_0\cap\rho(B_0)\ne\varnothing$.
Hence, the decreasing sets $A_0,~B_0$ are nonempty and contain $\mathbf{0}$,
which implies that $\boldsymbol\tau\notin D_0^A\cup D_0^B$.
Hence, $\prod_{j=1}^m(h_j-\alpha_j)\ge2$ for $\boldsymbol\alpha\in D_0^A\cup D_0^B$
and \eqref{eq:multivariate-stabilizer} yields $\sigma=\delta_0+1$.
Let $J=U\times\mathcal S_{m+1}$, and let $E_0$ and $F_0$ be as in Lemma~\ref{lem:block-relative-distance}
for either of its two cases. By Lemma~\ref{lem:block-relative-distance},
the corresponding relative minimum distance in \eqref{eq:general-local-relative-distance}
is $d(\mathscr C_{\mathcal S_{m+1}}(\mathbf E)\setminus\mathscr C_{\mathcal S_{m+1}}(\mathbf F))$
with $F_0\subseteq E_0$. If $F_0=E_0$, then this relative minimum distance and $d(E_0\setminus F_0)$
are both infinite. Otherwise, the hypotheses of Lemma~\ref{thm:coefficient-distance} hold,
and Lemma~\ref{thm:coefficient-distance}.2) with $s=1$ applies. Hence,
the two relative minimum distances in \eqref{eq:general-local-relative-distance} for
$J$ equal $\delta_0d(P_U(\C_{\mathcal X}(A_0))\setminus S_U(\C_{\mathcal X}(D_0^B)))$
and $\delta_0d(P_U(\C_{\mathcal X}(B_0))\setminus S_U(\C_{\mathcal X}(D_0^A)))$, respectively,
which gives \eqref{eq:s1-block-distance}.
\end{IEEEproof}

\section{Derived GKT-optimal impure qLRCs with explicit minimum localities}\label{sec:optimality}

Based on the results established in Section~\ref{sec:multivariate}, we now derive several families of
GKT-optimal impure qLRCs with explicit minimum localities from the CSS codes in \eqref{eq:gmcc-css}.
They provide positive answers to {\bf Problems}~\ref{prob:quantum-classical-locality} and
\ref{prob:disjoint-optimal-construction}
by specific choices of $m$, $s$, the evaluation sets, and the chains $\mathbf A$ and $\mathbf B$.
In particular, $\delta_0$ is fixed for the designed locally recoverable ability with $m$ and $s$
before the chains are chosen.
Once we obtain the CSS code $\mathcal Q$ in \eqref{eq:gmcc-css} with the designed $\delta_0$,
we first determine its minimum distance, impurity, and minimum localities for
{\em all admissible} $\delta\ge2$ (except for $\delta_0<\delta\le s+\delta_0-1$ when $s\ge2$),
and then decide its GKT-optimality using Lemma \ref{lem:blocks-recovering}.3).
For $m=1$, we also determine the values of $\delta$ for which
$\mathcal Q$ has $(r_{\mathcal Q}^{\min}(\delta),\delta)$-locality with $r_{\mathcal Q}^{\min}(\delta)<n-\delta+1$,
while neither $\C_X$ nor $\C_Z$ has classical $(r',\delta)$-locality for any $0<r'<n$.
More precisely, we use Theorem~\ref{thm:all-s1} in the following two ranges of $\delta$.
\begin{enumerate}[label=\arabic*),leftmargin=*,itemsep=0pt,topsep=1pt]
\item For $2\le\delta\le\delta_0$, the minimum localities and the sets attaining them
follow directly from Theorem~\ref{thm:all-s1}.1) once $\sigma=s+\delta_0$ is known.
However, the exact parameters, the minimum stabilizer weight,
and the impurity of $\mathcal Q$ require additional arguments.

\item For $s+\delta_0-1<\delta\le d$, we apply Theorem~\ref{thm:all-s1}.2) to obtain the minimum localities
and the sets attaining them. The key step is to determine $\lambda(U\times\mathcal S_{m+1})$
for nonempty $U\subseteq\mathcal X$, together with the integer $\gamma_0$ there.
\end{enumerate}

\subsection{Case I: $m=1$ with $A_i=\{0,\ldots,h_1-a_i\}$ and $B_i=\{0,\ldots,h_1-b_i\}$}\label{subsec:interval-chains}
In this subsection, we consider the case $m=1$ and
write $x$ for $\mathbf x$ and $J_x$ for $J_{\mathbf x}$.
Then $\Lambda=\{0,\ldots,h_1-1\}$ and every nonempty decreasing subset of $\Lambda$ has the form
$\{0,\ldots,h_1-c\}$ with $1\le c\le h_1$.
For integers $2\le a_0\le\cdots\le a_{s-1}\le h_1$ and $2\le b_0\le\cdots\le b_{s-1}\le h_1$,
we take
\begin{equation}\label{eq:profile-conditions}
 a_i+b_{s-1-i}\le h_1+1,~~A_i=\{0,\ldots,h_1-a_i\},~~{\rm and}~~B_i=\{0,\ldots,h_1-b_i\}~~{\rm for}~~0\le i<s,
\end{equation}
in \eqref{eq:gmcc-css}.
By \eqref{eq:dual-chains} and \eqref{eq:profile-conditions}, we have
\begin{equation}\label{eq:interval-duals}
 D_i^A=\{0,\ldots,a_{s-1-i}-2\},~~D_i^B=\{0,\ldots,b_{s-1-i}-2\}\subseteq A_i,~~{\rm and}~~T_i=\{b_{s-1-i}-1,\ldots,h_1-a_i\}\ne\varnothing
\end{equation}
for $0\le i<s$.
Hence, $\C_Z^\perp\subsetneq\C_X$ by Lemma~\ref{lem:gmcc-pair}.2).
To apply Lemma \ref{lem:blocks-recovering} and Theorem~\ref{thm:all-s1},
we need the parameters and impurity of the resulting CSS code $\mathcal Q$ in \eqref{eq:gmcc-css} for any $s$.

\begin{lemma}\label{thm:corrected-family}
The CSS code $\mathcal Q$ in \eqref{eq:gmcc-css} with the choices in \eqref{eq:profile-conditions} has parameters
$$\Dsb{h_1(s+2\delta_0-2),sh_1+2s-\sum_{i=0}^{s-1}(a_i+b_i),
\min\{a_i(s+\delta_0-1-i),~b_i(s+\delta_0-1-i):\,0\le i<s\}}_q.$$
Moreover, the following results hold.
\begin{enumerate}[label=\arabic*),leftmargin=*,itemsep=0pt,topsep=1pt,beginpenalty=10000]
  \item If $\mathcal Q$ has minimum distance $d>s+\delta_0$, then $\mathcal{Q}$ is
  impure and has minimum stabilizer weight $\sigma=s+\delta_0$.
  Otherwise, $\mathcal Q$ is pure and $\sigma\ge d$.

  \item If $a_i=a$ and $b_i=b$ for $0\le i<s$ with $a\le b$, then $\mathcal Q$ has parameters
$
 \dsb{h_1(s+2\delta_0-2),(h_1-a-b+2)s,a\delta_0}_q.
$
\end{enumerate}
\end{lemma}
\begin{IEEEproof}
Substituting $|A_i|=h_1-a_i+1$ and $|B_i|=h_1-b_i+1$ in
\eqref{eq:multivariate-logical-dimension} gives $\kappa=sh_1+2s-\sum_{i=0}^{s-1}(a_i+b_i)$.
Since $A_i=\{0,\ldots,h_1-a_i\}$, $\min\{h_1-\alpha:\,\alpha\in A_i\}=a_i$
is attained at the largest exponent $h_1-a_i$, which lies in $T_i$ by \eqref{eq:interval-duals}.
We then deduce from \eqref{eq:relative-lower-bounds} and \eqref{eq:monomial-relative-upper} that
$d_{XZ}=\min\{a_i(s+\delta_0-1-i):\,0\le i<s\}$.
Since the inequality $a_i+b_{s-1-i}\le h_1+1$ in \eqref{eq:profile-conditions}
is invariant under exchanging $a_i$ and $b_i$ for $0\le i<s$ together with $i\mapsto s-1-i$,
the formula for $d_{ZX}$ follows in the same way.
Hence, $d=\min\{d_{XZ},~d_{ZX}\}$ and $\mathcal{Q}$ has the described parameters.

1) By \eqref{eq:interval-duals}, the minima of $h_1-\alpha$ over $D_i^A$ and $D_i^B$ are
$h_1-a_{s-1-i}+2$ and $h_1-b_{s-1-i}+2$, respectively.
Using \eqref{eq:multivariate-stabilizer}, after reversing the index,
we have $\sigma=\min\bigl\{s+\delta_0,~\min\{(h_1-a_i+2)(\delta_0+i),~(h_1-b_i+2)(\delta_0+i):\,0\le i<s\}\bigr\}$.
By \eqref{eq:profile-conditions}, we have $h_1-a_i+2\ge b_{s-1-i}+1$. Since $\delta_0+i=s+\delta_0-1-(s-1-i)$,
we have $b_{s-1-i}(\delta_0+i)\ge d$ and
$(h_1-a_i+2)(\delta_0+i)\ge(b_{s-1-i}+1)(\delta_0+i)>b_{s-1-i}(\delta_0+i)\ge d$, and similarly for the terms involving $b_i$.
In summary, $\sigma$ is the minimum of $s+\delta_0$ and some terms that exceed $d$.
Hence, if $d>s+\delta_0$, then $\sigma=s+\delta_0<d$ and $\mathcal Q$ is impure.
Otherwise, $\sigma\ge d$ and $\mathcal Q$ is pure.

2) This follows by substituting $a_i=a$ and $b_i=b$ for $0\le i<s$ with $a\le b$
into the parameters of $\mathcal{Q}$.
\end{IEEEproof}
\medskip

When $\delta>s+\delta_0-1$, Theorem~\ref{thm:all-s1}.2) can reduce the minimum localities to
the sets $U\subseteq\mathcal X$ with $\lambda(U\times\mathcal S_2)\ge\delta$,
which determine the integer $\gamma_0$ there. For the CSS codes $\mathcal Q$ in Lemma~\ref{thm:corrected-family},
the following lemma determines the local relative distances $\lambda(U\times\mathcal S_2)$,
which depend only on $|U|$ and are nondecreasing in $|U|$.
\begin{lemma}\label{lem:interval-local-distance}
For the CSS code $\mathcal Q$ in \eqref{eq:gmcc-css} with the choices in \eqref{eq:profile-conditions} and every nonempty $U\subseteq\mathcal X$, we have
\begin{equation}\label{eq:interval-local-distance}
 \lambda(U\times\mathcal S_2)=\min\bigl\{(s+\delta_0-1-i)\max\{1,~|U|-h_1+c\}:\,0\le i<s,~c\in\{a_i,~b_i\}\bigr\}.
\end{equation}
In particular, the following results hold.
\begin{enumerate}[label=\arabic*),leftmargin=*,itemsep=0pt,topsep=1pt,beginpenalty=10000]
\item If $s=1$, $a_0=a$, and $b_0=b$ with $a\le b$, then
\begin{equation}\label{eq:recoverable-erasures}
 \lambda(U\times\mathcal S_2)=\delta_0\max\{1,~|U|-h_1+a\}.
\end{equation}
\item If $a_0=b_0=u$ and $a_i=b_i=u+1$ for $1\le i<s$, where $s\geq 2$, $u\ge2$ and $\delta_0\ge u(s-1)$, then
\begin{equation}\label{eq:two-factor-local-distance}
 \lambda(U\times\mathcal S_2)=
 \begin{cases}
 \delta_0,&|U|\le h_1-u,\\
 (s+\delta_0-1)(|U|-h_1+u),&h_1-u+1\le|U|\le h_1.
 \end{cases}
\end{equation}
\end{enumerate}
\end{lemma}
\begin{IEEEproof}
By \eqref{eq:general-local-relative-distance},
$\lambda(U\times\mathcal S_2)=
\min\{d(P_{U\times\mathcal S_2}(\C_X)\setminus S_{U\times\mathcal S_2}(\C_Z^\perp)),~
d(P_{U\times\mathcal S_2}(\C_Z)\setminus S_{U\times\mathcal S_2}(\C_X^\perp))\}$.
By Lemma~\ref{lem:block-relative-distance}.2), we have
$P_{U\times\mathcal S_2}(\C_X)=\mathscr C_{\mathcal S_2}(\mathbf E)$ and
$S_{U\times\mathcal S_2}(\C_Z^\perp)=\mathscr C_{\mathcal S_2}(\mathbf F)$
for the codes $E_i=P_U(\C_{\mathcal X}(A_i))$
and $F_i=S_U(\C_{\mathcal X}(D_i^B))$ of length $|U|$,
which consist of the $i$th restricted coefficients of the codewords of
$P_{U\times\mathcal S_2}(\C_X)$ and of $S_{U\times\mathcal S_2}(\C_Z^\perp)$, respectively.
By Lemma~\ref{lem:block-relative-distance}.1), we also have $E_0\supseteq\cdots\supseteq E_{s-1}$ and $F_i\subseteq E_i$.
For $0\le i<s$, the code $E_i$ consists of the evaluations on $U$ of
the polynomials in $\F_q[X_1]$ of degree at most $h_1-a_i$.
In terms of $|U|$, we have the following two cases for $d(E_i)$.
\begin{itemize}[leftmargin=*,itemsep=0pt,topsep=1pt]
\item If $|U|\ge h_1-a_i+1$, then a nonzero such polynomial has at most $h_1-a_i$ roots,
and hence this code is an $[|U|,h_1-a_i+1,|U|-h_1+a_i]_q$ MDS code.
\item If $|U|\le h_1-a_i$, then this code equals $\F_q^{|U|}$ by Lagrange interpolation,
and hence its minimum distance is one.
\end{itemize}

Combining these two cases, we derive $d(E_i)=\max\{1,~|U|-h_1+a_i\}\le a_i$.
In addition, \eqref{eq:interval-duals} implies that
the nonzero codewords of $F_i$ are evaluations on $U$ of
polynomials $\psi\in\F_q[X_1]$ with $\deg(\psi)\le b_{s-1-i}-2$ and $\psi(x)=0$ for $x\in\mathcal X\setminus U$.
Then these codewords have weight at least $|U|-(b_{s-1-i}-2-(h_1-|U|))=h_1-b_{s-1-i}+2\geq a_i+1$,
where the last inequality follows from \eqref{eq:profile-conditions}.
It follows that a codeword of minimum weight in $E_i$ lies outside $F_i$.
We then obtain that $F_i\ne E_i$ and
$d(E_i\setminus F_i)=d(E_i)=\max\{1,~|U|-h_1+a_i\}$.
Hence, $\mathbf E$ and $\mathbf F$ satisfy the hypotheses of
Lemma~\ref{thm:coefficient-distance}.3), which gives
\[
 d(P_{U\times\mathcal S_2}(\C_X)\setminus S_{U\times\mathcal S_2}(\C_Z^\perp))=\min\{(s+\delta_0-1-i)\max\{1,~|U|-h_1+a_i\}:\,0\le i<s\}.
\]
From \eqref{eq:profile-conditions}, we also have $h_1-a_{s-1-i}+2\ge b_i+1$.
Applying the same argument as above with $B_i$ and $D_i^A$ then provides the corresponding formula
for $d(P_{U\times\mathcal S_2}(\C_Z)\setminus S_{U\times\mathcal S_2}(\C_X^\perp))$ with $b_i$ in place of $a_i$.
This proves \eqref{eq:interval-local-distance}.

1) Since $a\le b$, the minimum in \eqref{eq:interval-local-distance} is attained at $c=a$.

2) Since $a_i=b_i$ for $0\le i<s$, we verify the following two cases.
\begin{itemize}
  \item If $|U|\le h_1-u$, then $|U|-h_1+a_i\le1$ for $0\le i<s$.
  Hence, every term in \eqref{eq:interval-local-distance} is $s+\delta_0-1-i$
  with the minimum $\delta_0$ attained by taking $i=s-1<s$.

  \item If $h_1-u+1\le|U|\le h_1$, then $1\le|U|-h_1+u\le u$.
  Since $\delta_0\ge u(s-1)\ge(|U|-h_1+u)(s-1)$,
  we deduce that
  $\min\{(s+\delta_0-1)(|U|-h_1+u),~\delta_0(|U|-h_1+u+1)\}=(s+\delta_0-1)(|U|-h_1+u)$.
\end{itemize}
We have completed the whole proof. 
\end{IEEEproof}
\bigskip 

\subsubsection{{\bf \em When $s=1$}}\label{subsec:m-s-1}

Combining Lemma~\ref{lem:interval-local-distance}.1) with Theorem~\ref{thm:all-s1},
we determine the parameters, the minimum localities, and the recovering sets of minimum size of
the corresponding CSS codes.

\begin{theorem}\label{cor:interval-recovery}
Let $s=1$, $\delta_0\ge2$, and $q\ge\max\{h_1,~2\delta_0-1\}$, and let $a_0=a$ and $b_0=b$ with $a\le b$.
Then the CSS code $\mathcal Q$ in \eqref{eq:gmcc-css} with the choices in \eqref{eq:profile-conditions}
is an impure $$\dsb{h_1(2\delta_0-1),h_1-a-b+2,a\delta_0}_q$$ qLRC with minimum locality
\begin{equation}\label{eq:complete-quantum-locality}
 r_{\mathcal Q}^{\min}(\delta)=
 \begin{cases}
 \delta_0,&2\le\delta\le\delta_0,\\
 (2\delta_0-1)(h_1-a+\lceil\delta/\delta_0\rceil)-\delta+1,&\delta_0<\delta\le a\delta_0.
 \end{cases}
\end{equation}
Moreover, the following results hold.
\begin{enumerate}[label=\arabic*),leftmargin=*,itemsep=0pt,topsep=1pt,beginpenalty=10000]
\item For $2\le\delta\le\delta_0$, $r_{\mathcal Q}^{\min}(\delta)$ is attained exactly by
the sets $J\subseteq J_x$ with $x\in\mathcal X$ and $|J|=\delta_0+\delta-1$, where $J_x$ is the block in \eqref{eq:disjoint-recovering-partition}.
\item For $\delta_0<\delta\le a\delta_0$, $r_{\mathcal Q}^{\min}(\delta)$ is attained exactly by the sets $U\times\mathcal S_2$ with $U\subseteq\mathcal X$ and $|U|=h_1-a+\lceil\delta/\delta_0\rceil$.
Moreover, these sets are proper if and only if $\delta\le(a-1)\delta_0$, and any two $(r,\delta)$-recovering sets intersect.
\item For $2\le\delta\le a\delta_0$, $\mathcal Q$ has $(r_{\mathcal Q}^{\min}(\delta),\delta)$-locality with $r_{\mathcal Q}^{\min}(\delta)<n-\delta+1$
while neither $\C_X$ nor $\C_Z$ has classical $(r',\delta)$-locality
for any $0<r'<n$ if and only if $\delta_0+2\le\delta\le(a-1)\delta_0$.
\end{enumerate}
\end{theorem}
\begin{IEEEproof}
By Corollary~\ref{cor:s1-specialization} and Lemma~\ref{thm:corrected-family}.2),
the CSS code $\mathcal Q$ has the stated parameters and $\sigma=\delta_0+1<a\delta_0=d$.
Hence, $\mathcal Q$ is impure.
By Theorem~\ref{lem:css-support-restrictions}.1), the admissible range for
its locality is $2\le\delta\le d$.
We prove \eqref{eq:complete-quantum-locality} together with 1) and 2).

1) The described result holds immediately by taking $s=1$ in Theorem~\ref{thm:all-s1}.1).

2) For $\delta_0<\delta\le a\delta_0$,
it follows from \eqref{eq:recoverable-erasures} that
$\lambda(U\times\mathcal S_2)\ge\delta$ if and only if $|U|\ge h_1-a+\lceil\delta/\delta_0\rceil$.
Since $\delta\le a\delta_0$, we have $h_1-a+\lceil\delta/\delta_0\rceil\le h_1$.
The facts $2\delta_0-1\ge\delta_0$ and $h_1-a+\lceil\delta/\delta_0\rceil\ge\lceil\delta/\delta_0\rceil$
then yield $(2\delta_0-1)(h_1-a+\lceil\delta/\delta_0\rceil)\ge\delta$.
Hence, applying Theorem~\ref{thm:all-s1}.2) with $\gamma_0=h_1-a+\lceil\delta/\delta_0\rceil$
gives \eqref{eq:complete-quantum-locality} for these $\delta$ and the sets attaining it.

Moreover, these sets are proper if and only if $h_1-a+\lceil\delta/\delta_0\rceil\le h_1-1$,
that is, $\delta\le(a-1)\delta_0$.
By Theorem~\ref{thm:all-s1}.2), every $(r,\delta)$-recovering set of $\mathcal Q$ is
$U\times\mathcal S_2$ with $|U|\ge h_1-a+\lceil\delta/\delta_0\rceil$.
Since $\delta>\delta_0$ and $2a\le a+b\le h_1+1$ by \eqref{eq:profile-conditions},
we further deduce that $|U|\ge h_1-a+2>h_1/2$.
Hence, any two such sets $U\times\mathcal S_2$ and $U'\times\mathcal S_2$ satisfy $|U|+|U'|>h_1=|\mathcal X|$,
which implies $U\cap U'\ne\varnothing$.

3) According to Lemma~\ref{lem:cartesian-duality-distance}.1),
$d(\C_X)=\min\{\delta_0+1,~a\delta_0\}=\delta_0+1$ and
$d(\C_Z)=\min\{\delta_0+1,~b\delta_0\}=\delta_0+1$.
Recall that any nontrivial linear code $\C$ has classical $(r',\delta)$-locality for some $0<r'<n$
if and only if $\delta\le d(\C)$.
Hence, neither $\C_X$ nor $\C_Z$ has classical $(r',\delta)$-locality for any $0<r'<n$
if and only if $\delta\ge\delta_0+2$.
In particular, both sides of 3) fail for $2\le\delta\le\delta_0+1$.
Now let $\delta_0+2\le\delta\le d$. Since $\delta_0<\delta\le a\delta_0$,
the sets in 2) have size $r_{\mathcal Q}^{\min}(\delta)+\delta-1$, and hence,
$r_{\mathcal Q}^{\min}(\delta)<n-\delta+1$ if and only if $\delta\le(a-1)\delta_0$ by 2).
We complete the proof by noticing that $\mathcal Q$ has
$(r_{\mathcal Q}^{\min}(\delta),\delta)$-locality by the definition of $r_{\mathcal Q}^{\min}(\delta)$.
\end{IEEEproof}
\medskip

Since any two $(r,\delta)$-recovering sets intersect for $\delta_0<\delta\le a\delta_0$
in Theorem~\ref{cor:interval-recovery}.2), we only consider the GKT-optimality of these CSS codes
for the disjoint $(\delta_0,\delta_0)$-recovering sets $J_x$ in the following theorem,
where further values of $a$ and $b$ are required.

\begin{theorem}\label{thm:optimal-attainment}
Let $h_1\ge3$, $\delta_0\ge2$, $1\le\kappa\le h_1-2$,
and $q\ge\max\{h_1,~2\delta_0-1\}$, and keep the notation of Theorem~\ref{cor:interval-recovery} with
$a=\lfloor(h_1-\kappa)/2\rfloor+1$ and $b=\lceil(h_1-\kappa)/2\rceil+1$.
Then the CSS code $\mathcal Q$ is a GKT-optimal
\[
 \dsb{h_1(2\delta_0-1),\kappa,
 \delta_0(\lfloor(h_1-\kappa)/2\rfloor+1)}_q
\]
impure $(\delta_0,\delta_0)$-qLRC with minimum locality
\begin{equation}\label{eq:optimal-attainment-locality}
 r_{\mathcal Q}^{\min}(\delta)=
 \begin{cases}
 \delta_0,&2\le\delta\le\delta_0,\\
 (2\delta_0-1)(\lceil(h_1+\kappa)/2\rceil+\lceil\delta/\delta_0\rceil-1)-\delta+1,&\delta_0<\delta\le\delta_0(\lfloor(h_1-\kappa)/2\rfloor+1).
 \end{cases}
\end{equation}
Moreover, if $a\ge3$, then for $\delta_0+2\le\delta\le(a-1)\delta_0$,
$\mathcal Q$ has $(r_{\mathcal Q}^{\min}(\delta),\delta)$-locality with
$r_{\mathcal Q}^{\min}(\delta)< n-\delta+1$ given by \eqref{eq:optimal-attainment-locality},
while neither $\C_X$ nor $\C_Z$ has classical $(r',\delta)$-locality for any $0<r'<n$.
\end{theorem}
\begin{IEEEproof}
By the choice of $a$ and $b$, we have $2\le a\le b\le a+1$ and $a+b=h_1-\kappa+2\le h_1+1$,
satisfying \eqref{eq:profile-conditions}.
After applying Theorem~\ref{cor:interval-recovery} with these $a$ and $b$,
it suffices to show that \eqref{eq:complete-quantum-locality} takes the form \eqref{eq:optimal-attainment-locality} and that $\mathcal Q$ is GKT-optimal.
Since $h_1-a=h_1-\lfloor(h_1-\kappa)/2\rfloor-1=\lceil(h_1+\kappa)/2\rceil-1$ and $a\delta_0=\delta_0(\lfloor(h_1-\kappa)/2\rfloor+1)$,
substituting them into \eqref{eq:complete-quantum-locality} gives \eqref{eq:optimal-attainment-locality}.
Since $b\le a+1$, we have $\kappa=h_1-a-b+2>h_1-2a$.
As $d=a\delta_0$, the GKT-optimality then follows from Lemma~\ref{lem:blocks-recovering}.3) with $c=a$.
\end{IEEEproof}
\bigskip

\subsubsection{{\bf \em When $s\ge2$}}\label{subsec:s-ge-2}

We first show that taking the constant chains with $a_i=b_i=a$ for $0\le i<s$ in \eqref{eq:profile-conditions}
{\em cannot} provide GKT-optimal impure qLRCs.
In this case, we obtain $\kappa=s(h_1-2a+2)$ and $d=a\delta_0$ by Lemma~\ref{thm:corrected-family}.2).
Since $a\delta_0=a(s+\delta_0-1)-a(s-1)$, it can be further checked that the difference between the right-hand side of \eqref{eq:gkt-blocks-optimal} and $\kappa$ is
\[
\Delta=2\bigl[s(a-1)-a\delta_0+(\delta_0-1)\lceil a\delta_0/(s+\delta_0-1)\rceil\bigr]=2\bigl[a(s-1)-s-(\delta_0-1)\lfloor a(s-1)/(s+\delta_0-1)\rfloor\bigr].
\]
We have the following cases.
\begin{itemize}
  \item
If $a(s-1)<s+\delta_0-1$, then $\lfloor a(s-1)/(s+\delta_0-1)\rfloor=0$, and $a\ge2$ gives $a(s-1)-s\ge s-2\ge0$,
implying $\Delta\geq 0$.

\item
Otherwise, $\lfloor a(s-1)/(s+\delta_0-1)\rfloor\ge1$, and $a(s-1)\ge (s+\delta_0-1)\lfloor a(s-1)/(s+\delta_0-1)\rfloor$ gives
$\Delta\geq 2s(\lfloor a(s-1)/(s+\delta_0-1)\rfloor-1)\ge0$.
\end{itemize}
In both cases, the right-hand side is at least $\kappa$.
These chains therefore do not give GKT-optimal codes for the disjoint $(s+\delta_0-1,\delta_0)$-recovering sets $J_x$.

By contrast, nonconstant integers $a_i$ and $b_i$ can give the following GKT-optimal impure qLRCs.

\begin{theorem}\label{thm:two-factor-optimality}
Let $s\geq 2$, $u\ge2$, $h_1\ge2u+1$, $\delta_0\ge u(s-1)$, and $q\ge\max\{h_1,~s+2\delta_0-2\}$,
and let $a_0=b_0=u$ and $a_i=b_i=u+1$ for $1\le i<s$.
Then the CSS code $\mathcal Q$ in \eqref{eq:gmcc-css} with the choices in \eqref{eq:profile-conditions} is a
GKT-optimal
$$\dsb{h_1(s+2\delta_0-2),s(h_1-2u)+2,u(s+\delta_0-1)}_q$$
impure $(s+\delta_0-1,\delta_0)$-qLRC with minimum locality
\begin{equation}\label{eq:nonconstant-high-delta-locality}
 r_{\mathcal Q}^{\min}(\delta)=
 \begin{cases}
 s+\delta_0-1,&2\le\delta\le\delta_0,\\
 (s+2\delta_0-2)(h_1-u+\lceil\delta/(s+\delta_0-1)\rceil)-\delta+1,&s+\delta_0-1<\delta\le u(s+\delta_0-1).
 \end{cases}
\end{equation}
Moreover, the following results hold.
\begin{enumerate}[label=\arabic*),leftmargin=*,itemsep=0pt,topsep=1pt,beginpenalty=10000]
\item For $2\le\delta\le\delta_0$, $r_{\mathcal Q}^{\min}(\delta)$ is attained exactly by
the sets $J\subseteq J_x$ with $x\in\mathcal X$ and $|J|=s+\delta_0+\delta-2$,
where $J_x$ is the block in \eqref{eq:disjoint-recovering-partition}.

\item For $s+\delta_0-1<\delta\le u(s+\delta_0-1)$, $r_{\mathcal Q}^{\min}(\delta)$ is attained exactly by
the sets $U\times\mathcal S_2$ with $U\subseteq\mathcal X$ and $|U|=h_1-u+\lceil\delta/(s+\delta_0-1)\rceil$.
Moreover, these sets are proper if and only if $\delta\le(u-1)(s+\delta_0-1)$,
and any two $(r,\delta)$-recovering sets intersect.

\item For $2\le\delta\le u(s+\delta_0-1)$, $\mathcal Q$ has $(r_{\mathcal Q}^{\min}(\delta),\delta)$-locality
with $r_{\mathcal Q}^{\min}(\delta)<n-\delta+1$ while neither $\C_X$ nor $\C_Z$ has
classical $(r',\delta)$-locality for any $0<r'<n$ if and only if
$s+\delta_0+1\le\delta\le(u-1)(s+\delta_0-1)$.
\end{enumerate}
\end{theorem}
\begin{IEEEproof}
Since $a_i+b_{s-1-i}\le2u+2\le h_1+1$, these integers satisfy the inequality in \eqref{eq:profile-conditions}.
Since $\sum_{i=0}^{s-1}(a_i+b_i)=2(su+s-1)$ and $(u+1)\delta_0-u(s+\delta_0-1)=\delta_0-u(s-1)\ge0$,
it follows from Lemma~\ref{thm:corrected-family} that $\mathcal{Q}$ has the stated parameters.
Since $u\ge2$, we have $u(s+\delta_0-1)>s+\delta_0$, and hence,
$\sigma=s+\delta_0$ and $\mathcal Q$ is impure by Lemma~\ref{thm:corrected-family}.1).
The GKT-optimality of $\mathcal Q$ directly follows from
taking $c=u$ in Lemma~\ref{lem:blocks-recovering}.3).

1) Since $\sigma=s+\delta_0$, Theorem~\ref{thm:all-s1}.1) gives the described \eqref{eq:nonconstant-high-delta-locality} for $2\le\delta\le\delta_0$ and the sets attaining it.
It remains to show that every $(s+\delta_0-1,\delta)$-recovering set $J$ of $\mathcal Q$ satisfies $J\subseteq J_x$ for some $x\in\mathcal X$.
As in the proof of Theorem~\ref{thm:all-s1}.1), we have $|J|=s+\delta_0+\delta-2\le s+2\delta_0-2$ and $d(P_J(\C_X))\ge\delta\ge2$.

Suppose that $J$ meets $k\ge2$ blocks, and choose $x\in\mathcal X$ such that $\mu=|J\cap J_x|$ is the smallest among the positive integers $|J\cap J_{\xi}|$ with $\xi\in\mathcal X$. Then $k\mu\le|J|$.
If $\mu\le\delta_0-1$, then the argument in the proof of Theorem~\ref{thm:all-s1}.1) for $s=1$, which only uses the lower coefficients, gives a codeword of $P_J(\C_X)$ of weight one, a contradiction.
Hence, $\mu\ge\delta_0$, and it follows from $\delta_0\ge u(s-1)\ge2s-2>s-2$ that $k\delta_0\le|J|\le s+2\delta_0-2<3\delta_0$,
yielding $k=2$.

Let $x'\in\mathcal X$ with $x'\ne x$ and $J\cap J_{x'}\ne\varnothing$, and fix $(x,z)\in J\cap J_x$.
As $\mu\le|J|/2\le s/2+\delta_0-1\le s+\delta_0-1$, we obtain a polynomial
$p=\sum_{\ell=0}^{s+\delta_0-2}p_\ell X_2^\ell\in\F_q[X_2]$ with $p(z)=1$ and $p(y)=0$
for every $(x,y)\in(J\cap J_x)\setminus\{(x,z)\}$ by Lagrange interpolation.
Let $\psi=(X_1-x')/(x-x')\in\F_q[X_1]$ and $\boldsymbol\psi=(\psi(\xi))_{\xi\in\mathcal X}$.
Since $\deg_{X_1}(\psi)=1\le u\le h_1-u-1\le h_1-a_i$, every exponent of $\psi$ lies in $A_i=\{0,\ldots,h_1-a_i\}$,
and hence $\boldsymbol\psi\in\C_{\mathcal X}(A_i)$ for $0\le i<s$.
Taking $\mathbf g_\ell=p_\ell\boldsymbol\psi$ and $\mathbf u_i=p_{\delta_0-1+i}\boldsymbol\psi$
in \eqref{eq:linear-coefficient-evaluation} with $U=\mathcal X$,
we then obtain a codeword of $\C_X$ with $f_x=p$ and $f_{x'}=0$,
whose restriction to $J$ has weight one.
This contradicts the fact $d(P_J(\C_X))\ge2$.
In summary, we deduce $k=1$ and $J\subseteq J_x$ for some $x\in\mathcal X$.

2) Let $s+\delta_0-1<\delta\le u(s+\delta_0-1)$ and $\varnothing\ne U\subseteq\mathcal X$.
By Lemma~\ref{lem:interval-local-distance}.2), $\lambda(U\times\mathcal S_2)$
is given by \eqref{eq:two-factor-local-distance}, which shows that
$\lambda(U\times\mathcal S_2)\ge\delta$ if and only if $|U|\ge h_1-u+\lceil\delta/(s+\delta_0-1)\rceil$.
Since $\delta\le u(s+\delta_0-1)$ and $h_1\geq u\geq 2$,
we obtain $h_1-u+\lceil\delta/(s+\delta_0-1)\rceil\le h_1$ and
$(s+2\delta_0-2)(h_1-u+\lceil\delta/(s+\delta_0-1)\rceil)\ge\delta$.
Hence, taking $\gamma_0=h_1-u+\lceil\delta/(s+\delta_0-1)\rceil$ in Theorem~\ref{thm:all-s1}.2)
then gives the described \eqref{eq:nonconstant-high-delta-locality} for these $\delta$ and the sets attaining it.

Moreover, it is easy to check that these sets are proper if and only if $h_1-u+\lceil\delta/(s+\delta_0-1)\rceil\le h_1-1$,
which is further equivalent to $\delta\le(u-1)(s+\delta_0-1)$.
Since $\delta>s+\delta_0-1$ and $h_1\ge2u+1$, we have $h_1-u+\lceil\delta/(s+\delta_0-1)\rceil\ge h_1-u+2>h_1/2$.
By Theorem~\ref{thm:all-s1}.2), every $(r,\delta)$-recovering set of $\mathcal Q$ is $U\times\mathcal S_2$ with
$|U|\ge h_1-u+\lceil\delta/(s+\delta_0-1)\rceil>h_1/2$.
This implies that any two such sets $U\times\mathcal S_2$ and $U'\times\mathcal S_2$ satisfy
$|U|+|U'|>h_1=|\mathcal X|$, and thereby $U\cap U'\ne\varnothing$.

3) By Lemma~\ref{lem:cartesian-duality-distance}.1) and the conditions given in this theorem,
since $\C_Z=\C(\mathbf A)\diag(\mathbf w)$,
we have $d(\C_X)=d(\C_Z)=\min\{s+\delta_0,~u(s+\delta_0-1),~(u+1)\delta_0\}=s+\delta_0$.
Recall again that any nontrivial linear code $\C$ has classical $(r',\delta)$-locality for some $0<r'<n$
if and only if $\delta\le d(\C)$. Hence, neither $\C_X$ nor $\C_Z$ has classical $(r',\delta)$-locality for
any $0<r'<n$ if and only if $\delta\ge s+\delta_0+1$.
In particular, both sides of 3) fail for $2\le\delta\le s+\delta_0$.
Now let $s+\delta_0+1\le\delta\le d$.
Combining with $\delta\le u(s+\delta_0-1)$,
we deduce that the sets in 2) have size $r_{\mathcal Q}^{\min}(\delta)+\delta-1$,
and hence, $r_{\mathcal Q}^{\min}(\delta)<n-\delta+1$ if and only if $\delta\le(u-1)(s+\delta_0-1)$ by 2),
where the left-side statement is equivalent to saying that 
$\mathcal Q$ has $(r_{\mathcal Q}^{\min}(\delta),\delta)$-locality. 
\end{IEEEproof}

\medskip

So far, Theorems~\ref{thm:optimal-attainment} and~\ref{thm:two-factor-optimality}
have already given two infinite families of GKT-optimal impure qLRCs,
providing positive answers to {\bf Problem}~\ref{prob:disjoint-optimal-construction}.
Since {\bf Problem}~\ref{prob:quantum-classical-locality} concerns
impure CSS codes of the form $\CSS(\C,\C)$, we now specialize the GKT-optimal impure qLRCs derived in
Theorems~\ref{thm:optimal-attainment} and~\ref{thm:two-factor-optimality} to the case where $\C_X=\C_Z$.

If $\mathcal S_1,\ldots,\mathcal S_{m+1}$ are respective cosets of additive subgroups
$V_1,\ldots,V_{m+1}$ of $\F_q$,
then $\{\xi_j-z:\,z\in\mathcal S_j\setminus\{\xi_j\}\}=V_j\setminus\{0\}$ for $\xi_j\in\mathcal S_j$.
In this case, \eqref{eq.w} yields a constant vector $\mathbf w$ and $\C_Z=\C(\mathbf B)\diag(\mathbf w)=\C(\mathbf B)$.
Since the evaluations of the monomials $\mathbf X^{\mathbf i}$ with $\mathbf i\in\mathcal H$ are linearly independent by Lemma~\ref{lem:cartesian-duality-distance}.1) with $\Omega=\mathcal H$,
it follows that $\C_Z=\C_X$ if and only if $\mathbf B=\mathbf A$,
which excludes the case of odd $h_1-\kappa$ in Theorem~\ref{thm:optimal-attainment}.
We immediately obtain the following corollary,
which gives positive answers to {\bf Problem}~\ref{prob:quantum-classical-locality}.

\begin{corollary}\label{cor:cx-cz}
Suppose that $\mathcal S_1,\ldots,\mathcal S_{m+1}$ are cosets of additive subgroups of $\F_q$.
Then the following results hold.
\begin{enumerate}[label=\arabic*),leftmargin=*,itemsep=0pt,topsep=1pt,beginpenalty=10000]
\item For even $h_1-\kappa$, the GKT-optimal impure qLRC $\mathcal Q=\CSS(\C(\mathbf A),\C(\mathbf A))$ in Theorem~\ref{thm:optimal-attainment}
has $(r_{\mathcal Q}^{\min}(\delta),\delta)$-locality with $r_{\mathcal Q}^{\min}(\delta)<n-\delta+1$ given by \eqref{eq:optimal-attainment-locality} for $\delta_0+2\le\delta\le(a-1)\delta_0$,
while $\C(\mathbf A)$ has no classical $(r',\delta)$-locality for any $0<r'<n$.

\item The GKT-optimal impure qLRC $\mathcal Q=\CSS(\C(\mathbf A),\C(\mathbf A))$ in
Theorem~\ref{thm:two-factor-optimality} has $(r_{\mathcal Q}^{\min}(\delta),\delta)$-locality
with $r_{\mathcal Q}^{\min}(\delta)<n-\delta+1$ given by \eqref{eq:nonconstant-high-delta-locality}
for $s+\delta_0+1\le\delta\le(u-1)(s+\delta_0-1)$,
while $\C(\mathbf A)$ has no classical $(r',\delta)$-locality for any $0<r'<n$.
\end{enumerate}
\end{corollary}

\subsection{Case II: $m\ge2$ and $s=1$ with $A_0=\Lambda\setminus\{\boldsymbol\tau\}$ and two $B_0$}\label{subsec:m-ge-2}
In this subsection, let $s=1$ and $m\ge2$, and let $\mathbf e_j$ denote the $j$th
{\em standard coordinate vector} of length $m$.
Recall from \eqref{eq:gmcc-notation} that $\Lambda=\prod_{j=1}^m\{0,\ldots,h_j-1\}$ and
that $\boldsymbol\tau=(h_1-1,\ldots,h_m-1)$ is the largest element of $\Lambda$ in the componentwise order,
with $\rho(\boldsymbol\alpha)=\boldsymbol\tau-\boldsymbol\alpha$ and $\boldsymbol\alpha\in\Lambda$.
When $s=1$, the required chains $\mathbf A$ and $\mathbf B$ then consist of single decreasing sets
$A_0,~B_0\subseteq\Lambda$, that is, $\mathbf A=(A_0)$ and $\mathbf B=(B_0)$.
In \eqref{eq:gmcc-css}, take either
\begin{equation}\label{eq:multivariate-choice-1}
 A_0=B_0=\Lambda\setminus\{\boldsymbol\tau\}, ~~{\rm or}
\end{equation}
\begin{equation}\label{eq:multivariate-choice-2}
 A_0=\Lambda\setminus\{\boldsymbol\tau\}~~{\rm and}~~
 B_0=\Lambda\setminus\{\boldsymbol\tau,~\boldsymbol\tau-\mathbf e_k\}~~{\rm for~some}~~ 1\le k\le m.
\end{equation}
By \eqref{eq:dual-chains}, we have $D_0^A=\{\mathbf 0\}$, and $D_0^B=\{\mathbf 0\}$
or $D_0^B=\{\mathbf 0,~\mathbf e_k\}$ for \eqref{eq:multivariate-choice-1} and \eqref{eq:multivariate-choice-2},
respectively.
Since $m\ge2$ and $h_j\ge2$ for $1\le j\le m$, the coordinates of
$\boldsymbol\tau$ and of $\boldsymbol\tau-\mathbf e_k$
at every index $j\neq k$ with $1\le j\le m$ are equal to $h_j-1\ge1$.
Hence, $\boldsymbol\tau,~\boldsymbol\tau-\mathbf e_k\notin\{\mathbf 0,~\mathbf e_k\}$,
indicating that $D_0^B\subseteq A_0$ and $D_0^A\subseteq B_0$ for both choices.

Moreover, since $|\Lambda|=\prod_{j=1}^mh_j\ge4$,
we have $|A_0|\ge3>|D_0^B|$ and $|B_0|\ge2>|D_0^A|$, and hence both of the above inclusions are strict.
With Lemma~\ref{lem:gmcc-pair}.2), we know that $\C_Z^\perp\subsetneq\C_X$ under such settings.
In the next two theorems, we derive two infinite families of GKT-optimal impure qLRCs with
explicit minimum localities.

\begin{theorem}\label{thm:multivariate-s1}
Let $s=1$, $m\ge2$, $h_j\ge2$ for $1\le j\le m$, $\delta_0\ge2$,
and $q\ge\max\{h_1,\ldots,h_m,~2\delta_0-1\}$.
Then the CSS code $\mathcal Q$ in \eqref{eq:gmcc-css} with the choices in \eqref{eq:multivariate-choice-1}
is a GKT-optimal
$$\Dsb{(2\delta_0-1)\prod_{j=1}^mh_j,\prod_{j=1}^mh_j-2,2\delta_0}_q$$
impure $(\delta_0,\delta_0)$-qLRC with minimum locality
\begin{equation}\label{eq:multivariate-locality}
 r_{\mathcal Q}^{\min}(\delta)=
 \begin{cases}
 \delta_0,&2\le\delta\le\delta_0,\\
 (2\delta_0-1)\prod_{j=1}^mh_j-\delta+1,&\delta_0<\delta\le2\delta_0.
 \end{cases}
\end{equation}
Moreover, the following results hold.
\begin{enumerate}[label=\arabic*),leftmargin=*,itemsep=0pt,topsep=1pt,beginpenalty=10000]
\item For $2\le\delta\le\delta_0$, $r_{\mathcal Q}^{\min}(\delta)$ is attained exactly
by the sets $J\subseteq J_{\mathbf x}$ with $\mathbf x\in\mathcal X$ and $|J|=\delta_0+\delta-1$,
where $J_{\mathbf x}$ is the block in \eqref{eq:disjoint-recovering-partition}.

\item For $\delta_0<\delta\le2\delta_0$, $r_{\mathcal Q}^{\min}(\delta)$ is attained exactly
by the full coordinate set $\mathcal P$, which is the only $(r,\delta)$-recovering set of $\mathcal Q$.
\end{enumerate}
\end{theorem}
\begin{IEEEproof}
Based on Lemma~\ref{lem:gmcc-pair}.3), we obtain $\kappa=|A_0|+|B_0|-\prod_{j=1}^mh_j=\prod_{j=1}^mh_j-2$.
Since $\boldsymbol\tau\notin A_0\cup B_0$, every $\boldsymbol\alpha\in A_0\cup B_0$
satisfies $\prod_{j=1}^m(h_j-\alpha_j)\ge2$.
Then the lower bounds in \eqref{eq:relative-lower-bounds} give both $d_{XZ},~d_{ZX}\ge2\delta_0$.
With the conditions $m\ge2$ and $h_1,~h_2\ge2$, neither $\boldsymbol\tau-\mathbf e_1$ nor $\mathbf e_2$
equals $\mathbf 0$ or $\boldsymbol\tau$. Then $\boldsymbol\tau-\mathbf e_1$ and $\mathbf e_2$
lie in $T_0=A_0\setminus D_0^B=\Lambda\setminus\{\mathbf 0,~\boldsymbol\tau\}$,
and $\prod_{j=1}^m(h_j-(\boldsymbol\tau-\mathbf e_1)_j)=\prod_{j=1}^m((\mathbf e_2)_j+1)=2$.
Consequently, the upper bounds in \eqref{eq:monomial-relative-upper} are $2\delta_0$.
In summary, we get $d=2\delta_0$.
In addition, it follows from Corollary~\ref{cor:s1-specialization} that $\sigma=\delta_0+1<d$
and $\mathcal Q$ is impure.
Finally, with $s=1$, $d=2\delta_0$, and $\kappa=\prod_{j=1}^mh_j-2>\prod_{j=1}^mh_j-4$,
applying Lemma~\ref{lem:blocks-recovering}.3) with $c=2$ shows that $\mathcal Q$ is GKT-optimal.

1) The described result holds immediately by taking $s=1$ in Theorem~\ref{thm:all-s1}.1).

2)
Since $D_0^B=\{\mathbf{0}\}$,
$\C_{\mathcal X}(A_0)$ is a $[\prod_{j=1}^mh_j,\prod_{j=1}^mh_j-1,2]_q$ MDS code and
$d(\C_{\mathcal X}(D_0^B))=\prod_{j=1}^mh_j\ge2$
by Lemma~\ref{lem:cartesian-duality-distance}.1).
It follows that
$P_U(\C_{\mathcal X}(A_0))=\F_q^{|U|}$ for every $U\subsetneq\mathcal X$
and $S_U(\C_{\mathcal X}(D_0^B))$ contains no codeword of weight one, respectively.
Then $d(P_U(\C_{\mathcal X}(A_0))\setminus S_U(\C_{\mathcal X}(D_0^B)))=1$
for every $\varnothing\ne U\subsetneq\mathcal X$, and further
$\lambda(U\times\mathcal S_{m+1})\le\delta_0<\delta$ by applying it in \eqref{eq:s1-block-distance}.
However, $\lambda(\mathcal P)=2\delta_0\ge\delta$.
Therefore, for every $\varnothing\ne U\subseteq\mathcal X$,
$\lambda(U\times\mathcal S_{m+1})\ge\delta$ if and only if $U=\mathcal X$.

On the one hand, no set $U\times\mathcal S_{m+1}$ with
$\varnothing\ne U\subsetneq\mathcal X$ can be an $(r,\delta)$-recovering set of $\mathcal Q$
by Theorem~\ref{lem:css-support-restrictions}.
On the other hand, since $\delta>\delta_0=s+\delta_0-1$ for $s=1$,
we deduce from Theorem~\ref{thm:all-s1}.2) with $\gamma_0=|\mathcal X|$,
where $(2\delta_0-1)\gamma_0=n\ge d\ge\delta$,
that every $(r,\delta)$-recovering set coincides with $\mathcal X\times\mathcal S_{m+1}=\mathcal P$
and that $r_{\mathcal Q}^{\min}(\delta)=n-\delta+1=(2\delta_0-1)\prod_{j=1}^mh_j-\delta+1$,
which gives the second equation in \eqref{eq:multivariate-locality} and the only recovering set $\mathcal P$.
\end{IEEEproof}
\medskip

\begin{theorem}\label{thm:multivariate-s1-b}
Let $s=1$, $m\ge2$, $h_j\ge2$ for $1\le j\le m$, $\delta_0\ge2$, and $q\ge\max\{h_1,\ldots,h_m,~2\delta_0-1\}$.
Then the CSS code $\mathcal Q$ in \eqref{eq:gmcc-css} with the choices in \eqref{eq:multivariate-choice-2}
is a GKT-optimal
$$\Dsb{(2\delta_0-1)\prod_{j=1}^mh_j,\prod_{j=1}^mh_j-3,2\delta_0}_q$$
impure $(\delta_0,\delta_0)$-qLRC with the same minimum locality given by \eqref{eq:multivariate-locality}.
Moreover, the results 1) and~2) of Theorem~\ref{thm:multivariate-s1} also hold.
\end{theorem}
\begin{IEEEproof}
The proof is similar to that of Theorem~\ref{thm:multivariate-s1}, and we only give the differences.
Applying Lemma~\ref{lem:gmcc-pair}.3), we obtain $\kappa=|A_0|+|B_0|-\prod_{j=1}^mh_j=\prod_{j=1}^mh_j-3>\prod_{j=1}^mh_j-4$.
Since $m\ge2$, we can choose $k'\ne k$.
Then $\boldsymbol\tau-\mathbf e_k$ and $\mathbf e_{k'}$, in place of $\boldsymbol\tau-\mathbf e_1$ and $\mathbf e_2$, lie in $T_0=A_0\setminus D_0^B=\Lambda\setminus\{\mathbf 0,~\mathbf e_k,~\boldsymbol\tau\}$
and $\prod_{j=1}^m(h_j-(\boldsymbol\tau-\mathbf e_k)_j)=\prod_{j=1}^m((\mathbf e_{k'})_j+1)=2$.
In 2), we have $D_0^B=\{\mathbf 0,~\mathbf e_k\}$, and Lemma~\ref{lem:cartesian-duality-distance}.1) gives
$d(\C_{\mathcal X}(D_0^B))=(h_k-1)\prod_{j\ne k}h_j\ge|\Lambda|/2\ge2$.
\end{IEEEproof}
\medskip

\begin{remark}\label{rem:m2-css-cc}
  The two families of CSS codes obtained in Theorems~\ref{thm:multivariate-s1} and~\ref{thm:multivariate-s1-b}
  are GKT-optimal impure qLRCs with explicit minimum localities, and hence they also answer
  {\bf Problem}~\ref{prob:disjoint-optimal-construction}.
However, they do not give answers to {\bf Problem}~\ref{prob:quantum-classical-locality}, for the following reasons.
\begin{itemize}
  \item For $2\le\delta\le\delta_0$, we have $\delta<\delta_0+1=\sigma$ by Corollary~\ref{cor:s1-specialization}.
  Then Corollary~\ref{cor:locality-local-distance}.2) shows that the quantum $(r,\delta)$-recovering sets of these codes are exactly the common classical $(r,\delta)$-recovering sets of $\C_X$ and $\C_Z$.
  Hence, $\C_X$ and $\C_Z$ also have classical $(r_{\mathcal Q}^{\min}(\delta),\delta)$-locality.

  \item For $\delta_0<\delta\le2\delta_0$, \eqref{eq:multivariate-locality} gives $r_{\mathcal Q}^{\min}(\delta)=n-\delta+1$,
  and by 2) of Theorem~\ref{thm:multivariate-s1}, which also holds for Theorem~\ref{thm:multivariate-s1-b}, the only recovering set is the full coordinate set $\mathcal P$.
  Although $\C_X$ and $\C_Z$ even have no classical $(r',\delta)$-locality for any $0<r'<n$ when $\delta_0+2\le\delta\le2\delta_0$, since $d(\C_X)=d(\C_Z)=\delta_0+1$ by Lemma~\ref{lem:cartesian-duality-distance}.1),
  such a trivial locality is excluded in {\bf Problem}~\ref{prob:quantum-classical-locality}, which requires $r<n-\delta+1$.
\end{itemize}
\end{remark}

\subsection{Comparison and monotonicity of minimum localities}\label{subsec:cx-cz}

Recall that several infinite families of impure qLRCs have been constructed in \cite{GHMM2026QIP} and \cite{MS2026}.
Apart from special cases such as \cite[Theorems~10, 12, and~13]{GHMM2026QIP} and \cite[Proposition~32]{MS2026},
they form two explicit classes in \cite[Proposition~6]{GHMM2026QIP} and \cite[Theorem~28]{MS2026}.
In fact, these parameters can also be deduced from our construction as follows.
Some further advantages of our construction are also discussed.

\begin{enumerate}[label=\arabic*),leftmargin=*,itemsep=0pt,topsep=1pt]
\item The CSS code has parameters
$$\Dsb{HV,(2a'+(H\bmod2))(2b'+(V\bmod2)),\lceil(H+1)/2-a'\rceil\lceil(V+1)/2-b'\rceil}_q$$
and locality
$(V-\lceil(V-1)/2-b'\rceil,~\lceil(V+1)/2-b'\rceil)$ in \cite[Proposition~6]{GHMM2026QIP}
for integers $H,~V\ge3$, $(H-1)\bmod2\le a'<(H-1)/2$, and $(V-1)\bmod2\le b'<(V-1)/2$,
where $(H-1)\mid q-1$ and $(V-1)\mid q-1$ are required,
and it is impure if and only if the Equation (15) in the same paper holds.
These ranges of $a'$ and $b'$ give $\lceil(H+1)/2-a'\rceil\ge2$ and $\lceil(V+1)/2-b'\rceil\ge2$.
Since the dimension is positive, we have $2b'+(V\bmod2)\ge1$.
Then either $V$ is odd and $b'=0$, or $b'\ge1$.
Therefore, the complete parameters and locality can be revisited as follows.
\begin{itemize}[leftmargin=*,itemsep=0pt,topsep=1pt]
\item If $V$ is odd and $b'=0$, these parameters can be obtained by taking $h_1=H$,
$\delta_0=(V+1)/2$, and $\kappa=2a'+(H\bmod2)$ in Theorem~\ref{thm:optimal-attainment}.
In this case, our qLRC has locality $(\delta_0,\delta_0)$,
and its impurity, GKT-optimality, and minimum locality are further determined by Theorem~\ref{thm:optimal-attainment}.
\item If $b'\ge1$, these parameters can be obtained by taking $m=1$, $h_1=H$, $s=2b'+(V\bmod2)\ge2$,
$\delta_0=\lceil(V+1)/2-b'\rceil$, and the constant chains $a_i=b_i=\lceil(H+1)/2-a'\rceil$ for $0\le i<s$
at the beginning of Section~\ref{subsec:s-ge-2}. 
However, such parameters are further shown to be {\em not} GKT-optimal there.
\end{itemize}

\item The impure concatenated CSS code has parameters
$$\Dsb{h(2\delta-1),h-2d_{\rm out}+2,\delta d_{\rm out}}_q$$
and $(\delta,\delta)$-locality in \cite[Theorem~28]{MS2026}
for integers $d_{\rm out}\ge2$ and $h\ge2d_{\rm out}-1$,
where $q\ge\max\{h,~2\delta-1\}$ is required.
These parameters can be obtained by taking $h_1=h$, $\delta_0=\delta$, and $\kappa=h-2d_{\rm out}+2$ in Theorem~\ref{thm:optimal-attainment},
where $a=b=d_{\rm out}$ and the condition on $q$ is the same.
In this case, our qLRC has locality $(\delta_0,\delta_0)$,
and its GKT-optimality and minimum locality are further determined by Theorem~\ref{thm:optimal-attainment}.

\item Even for the parameters in 1) and 2), our constructions are more flexible
in the following respects.
\begin{itemize}[leftmargin=*,itemsep=0pt,topsep=1pt]

\item The qLRCs in \cite[Proposition~6]{GHMM2026QIP} require $(H-1)\mid q-1$ and $(V-1)\mid q-1$,
whereas ours only require $q\ge\max\{H,~V\}$.

\item Both \cite[Theorem~28]{MS2026} and the first case of 1) contain only even $h_1-\kappa$,
whereas every $1\le\kappa\le h_1-2$ is covered by Theorem~\ref{thm:optimal-attainment}
and odd $h_1-\kappa$ occurs when $b=a+1$.

\item For $d_{\rm out}=2$ and $m\ge2$, the same parameters with $h=\prod_{j=1}^mh_j$ arise
in Theorem~\ref{thm:multivariate-s1} under $q\ge\max\{h_1,\ldots,h_m,~2\delta_0-1\}$.
As $h$ may exceed $q$, the qLRCs in Theorem~\ref{thm:multivariate-s1} can have unbounded lengths over a fixed field,
whereas \cite{MS2026} requires $q\ge\max\{h,~2\delta-1\}$ and the length $n=h(2\delta-1)\le q^2$.
\end{itemize}

\item Moreover, our GKT-optimal impure $(r,\delta)$-qLRCs in
Theorems~\ref{thm:two-factor-optimality} and \ref{thm:multivariate-s1-b} can have parameter patterns
different from both known classes.

\begin{itemize}[leftmargin=*,itemsep=0pt,topsep=1pt]
\item For $s\ge2$, the nonconstant chains in Theorem~\ref{thm:two-factor-optimality} give GKT-optimal codes,
unlike the constant chains in the second case of 1).
For given $n$, $r$, and $\delta_0$, our qLRCs also have dimensions $\kappa\equiv sh_1+2\pmod{2s}$,
whereas $\kappa=s(h_1-2a+2)\equiv sh_1\pmod{2s}$ is required in the second case of 1).
In particular, these qLRCs have locality $(r,\delta)=(s+\delta_0-1,\delta_0)$ with $r\ne\delta$,
which further excludes 2).

\item The $(\delta_0,\delta_0)$-qLRCs in Theorem~\ref{thm:multivariate-s1-b} satisfy $n/(2\delta-1)-\kappa=3$.
However, both known classes of $(r,\delta)$-qLRCs have even $n/(2\delta-1)-\kappa$ when $r=\delta$.
\end{itemize}

\end{enumerate}

In summary, our constructions revisit the parameters of the explicit impure qLRCs in \cite{GHMM2026QIP} and \cite{MS2026},
and they also provide new GKT-optimal impure qLRCs with explicit minimum localities and more flexible parameters.

\begin{remark}\label{rem:nonmonotone-locality}
It is also interesting to observe that the minimum localities in
\eqref{eq:complete-quantum-locality}, \eqref{eq:nonconstant-high-delta-locality}, and
\eqref{eq:multivariate-locality} are not monotone in $\delta$.
\begin{itemize}[leftmargin=*,itemsep=0pt,topsep=1pt]
\item In \eqref{eq:complete-quantum-locality}, for $\delta_0<\delta<a\delta_0$,
we have $r_{\mathcal Q}^{\min}(\delta+1)=r_{\mathcal Q}^{\min}(\delta)-1$ if $\delta_0\nmid\delta$;
and $r_{\mathcal Q}^{\min}(\delta+1)=r_{\mathcal Q}^{\min}(\delta)+2\delta_0-2$ otherwise.
In addition, the same holds for \eqref{eq:nonconstant-high-delta-locality} and $s+\delta_0-1<\delta<u(s+\delta_0-1)$,
with $\delta_0$ and $2\delta_0-2$ replaced by $s+\delta_0-1$ and $s+2\delta_0-3$, respectively.

\item In \eqref{eq:multivariate-locality}, the minimum locality jumps from $\delta_0$ to $n-\delta_0$
at $\delta=\delta_0+1$ and then decreases by one at each step up to $\delta=2\delta_0$.
\end{itemize}

Table~\ref{tab:nonmonotone} lists some specific examples of $r_{\mathcal Q}^{\min}(\delta)$ for
GKT-optimal impure CSS codes from Theorems~\ref{thm:optimal-attainment}, \ref{thm:two-factor-optimality},
\ref{thm:multivariate-s1}, and~\ref{thm:multivariate-s1-b}.
In this table, boldface marks the values with $r_{\mathcal Q}^{\min}(\delta)<r_{\mathcal Q}^{\min}(\delta-1)$
and a dash means that $\delta>d$, which is impossible for the corresponding qLRCs by
Theorem~\ref{lem:css-support-restrictions}.1).
In particular, $r_{\mathcal Q}^{\min}(3)=13$ in the second row
is obtained by a computer verification, where $\delta=3$ is the only case that is not covered by Theorem~\ref{thm:two-factor-optimality}.

\begin{table}[h]
\centering
\caption{Examples of minimum localities $r_{\mathcal Q}^{\min}(\delta)$ of GKT-optimal impure CSS codes from Theorems~\ref{thm:optimal-attainment}, \ref{thm:two-factor-optimality}, \ref{thm:multivariate-s1}, and~\ref{thm:multivariate-s1-b}}\label{tab:nonmonotone}
\footnotesize
\setlength{\tabcolsep}{4pt}
\renewcommand{\arraystretch}{1.3}
\begin{tabular}{c|c|ccccc|c}
\toprule
\multirow{2}{*}{Choices} & \multirow{2}{*}{CSS codes} & \multicolumn{5}{c|}{$r_{\mathcal Q}^{\min}(\delta)$} & \multirow{2}{*}{Reference}\\
\cline{3-7}
 & & $\delta=2$ & $\delta=3$ & $\delta=4$ & $\delta=5$ & $\delta=6$ & \\
\midrule
$q=9$, $\mathcal S_1=\F_9$, $\mathcal S_2=\F_3\subseteq \F_9$, $\delta_0=2$, $a=b=3$ & $\dsb{27,5,6}_9$ & $2$ & $22$ & $\mathbf{21}$ & $23$ & $\mathbf{22}$ & Theorem~\ref{thm:optimal-attainment}\\
$q=5$, $\mathcal S_1=\F_5$, $\mathcal S_2=\{0,1,2,3\}$, $s=u=\delta_0=2$ & $\dsb{20,4,6}_5$ & $3$ & $13$ & $17$ & $\mathbf{16}$ & $\mathbf{15}$ & Theorem~\ref{thm:two-factor-optimality}\\
$q=3$, $\mathcal S_1=\mathcal S_2=\mathcal S_3=\F_3$, $\delta_0=2$ & $\dsb{27,7,4}_3$ & $2$ & $25$ & $\mathbf{24}$ & -- & -- & Theorem~\ref{thm:multivariate-s1}\\
$q=3$, $\mathcal S_1=\mathcal S_2=\mathcal S_3=\F_3$, $\delta_0=2$, $k=1$ & $\dsb{27,6,4}_3$ & $2$ & $25$ & $\mathbf{24}$ & -- & -- & Theorem~\ref{thm:multivariate-s1-b}\\
\bottomrule
\end{tabular}
\end{table}

\end{remark}

\section{Concluding remarks}\label{sec:conclusion}
This paper addressed {\bf Problem}~\ref{prob:quantum-classical-locality} and {\bf Problem}~\ref{prob:disjoint-optimal-construction}
by CSS codes constructed from GMCCs with decreasing defining sets.
Grouping the monomials by their degree in the last variable, we bounded the relative minimum distances of punctured codes with respect to shortened codes
through $s$ nested pairs of linear codes in the remaining variables in Lemmas~\ref{thm:coefficient-distance} and~\ref{lem:block-relative-distance}.
These bounds control both the minimum distance in Theorem~\ref{cor:gmcc-distance-specialization} and
the local relative distances that determine the minimum locality in Theorem~\ref{thm:all-s1},
and they are exact for the chains in Lemma~\ref{thm:corrected-family}.
For $m=s=1$, for $m=1$ and $s\ge2$, and for $m\ge2$ and $s=1$, GKT-optimal impure qLRCs are given in
Theorem~\ref{thm:optimal-attainment}, Theorem~\ref{thm:two-factor-optimality}, and Theorems~\ref{thm:multivariate-s1} and~\ref{thm:multivariate-s1-b}, respectively.
Their minimum localities are determined in Theorems~\ref{cor:interval-recovery},
\ref{thm:multivariate-s1}, and~\ref{thm:multivariate-s1-b}, and in Theorem~\ref{thm:two-factor-optimality}
except for $\delta_0<\delta\le s+\delta_0-1$,
which provide positive answers to {\bf Problem}~\ref{prob:disjoint-optimal-construction}.
In particular, the two families of GKT-optimal impure CSS codes $\mathcal Q=\CSS(\C(\mathbf A),\C(\mathbf A))$
in Corollary~\ref{cor:cx-cz} have $(r_{\mathcal Q}^{\min}(\delta),\delta)$-locality with
$r_{\mathcal Q}^{\min}(\delta)<n-\delta+1$ for values of $\delta$ at which
$\C(\mathbf A)$ has no classical $(r',\delta)$-locality for any $0<r'<n$,
which answers {\bf Problem}~\ref{prob:quantum-classical-locality} affirmatively.
In obtaining these results, we also
characterized the recovering sets of CSS codes by local relative distances,
obtained bounds on the relative minimum distances of CSS codes from GMCCs that can strictly exceed the known estimates,
determined the structure of the recovering sets of our qLRCs, and
found that their minimum localities need not be monotone in $\delta$.
We refer the reader to Section~\ref{subsec:contributions} for more details.

Note also that GMCCs are defined for arbitrary defining sets in Definition~\ref{def:cartesian-code},
whereas the CSS codes in this paper come only from the GMCCs with nested decreasing chains in
Definition~\ref{def:chain-gmcc}, whose defining sets $\Omega(\mathbf A)$
contain all monomials of degree at most $\delta_0-2$ in the distinguished variable $X_{m+1}$.
It would be worthwhile to investigate further whether the bounds in Section~\ref{sec:multivariate}
can be extended to CSS codes from general GMCCs,
and whether they can give further GKT-optimal impure qLRCs.

\section*{Acknowledgment}
The authors would like to thank Prof.~Markus Grassl for helpful discussions
on impure quantum locally recoverable codes during his visit to Nanyang
Technological University in August, 2026. During the preparation of this manuscript, 
the authors used Claude Opus~5.5
(Anthropic) to assist with the literature survey, language polishing,
and computational verification of the examples.
The authors take full responsibility for all statements, proofs, and claims of originality in this paper.

\end{document}